# Thickness-dependent self-polarization-induced intrinsic magnetoelectric effects in La$_{0.67}$Sr$_{0.33}$MnO$_3$/PbZr$_{0.52}$Ti$_{0.48}$O$_3$ heterostructures

Xiaoqian Yu[1], Lijun Wu[2], Bangmin Zhang[1], Hua Zhou[3], Yongqi Dong[3], Xiaohan Wu[1], Ronghui Kou[3], Ping Yang[4], Jingsheng Chen[1], Cheng-Jun Sun[3,*], Yimei Zhu[2,*], and Gan Moog Chow[1,*]

[1]*Department of Materials Science & Engineering, National University of Singapore, 117576, Singapore.*

[2]*Condensed Matter Physics & Materials Science Division, Brookhaven National Laboratory, New York 11973, USA.*

[3]*Advanced Photon Source, Argonne National Laboratory, Argonne, Illinois 60439, USA.*

[4]*Singapore Synchrotron Light Source, National University of Singapore, 117603, Singapore.*

*\*Corresponding Authors' Email: cjsun@aps.anl.gov; zhu@bnl.gov; msecgm@nus.edu.sg.*

# ABSTRACT

The self-polarization of PbZr$_{0.52}$Ti$_{0.48}$O$_3$ thin film is switched by changing film thickness through the competition between the strain relaxation-induced flexoelectric fields and the interfacial effects. Without an applied electric field, this reversal of self-polarization is exploited to control the magnetic properties of La$_{0.67}$Sr$_{0.33}$MnO$_3$ by the competition/cooperation between the charge-mediated and the strain-mediated effects. Scanning transmission electron microscopy, polarized near edge x-ray absorption spectroscopy, and half-integer diffraction measurements are employed to decode this intrinsic magnetoelectric effects in La$_{0.67}$Sr$_{0.33}$MnO$_3$/PbZr$_{0.52}$Ti$_{0.48}$O$_3$ heterostructures. With PbZr$_{0.52}$Ti$_{0.48}$O$_3$ films < 48 nm, the self-polarization-driven carrier density modulation around La$_{0.67}$Sr$_{0.33}$MnO$_3$/PbZr$_{0.52}$Ti$_{0.48}$O$_3$ interface and the strain-mediated Mn 3$d$ orbital occupancy work together to enhance magnetism of 14 unit cells La$_{0.67}$Sr$_{0.33}$MnO$_3$ film; with PbZr$_{0.52}$Ti$_{0.48}$O$_3$ layers > 48 nm, the strain-induced the change of bond length/angle of MnO$_6$ accompanied with a modified spin configuration are responsible for the decrease in Curie temperature and magnetization of 14 unit cells La$_{0.67}$Sr$_{0.33}$MnO$_3$ film.

# I. INTRODUCTION

Artificially designed multiferroic heterostructures of ferromagnetic (FM) /ferroelectric (FE) layers draw significant interests due to the rich physics of magnetoelectric (ME) coupling and potential applications [1-6]. Specially, the application of electric field to the FE layer in multiferroics to control magnetism in magnetic memory storage, sensors, and spintronics is of particular interests because of reduced power consumption [7]. For example, the reversal of FE polarization-induced electrostatic modulation in charge carrier density at $La_{0.8}Sr_{0.2}MnO_3/PbZr_{0.2}Ti_{0.8}O_3$ interface [8, 9], or interfacial orbital reconstruction in $La_{0.5}Sr_{0.5}MnO_3/BaTiO_3$ [10] and $La_{0.825}Sr_{0.175}MnO_3/PbZr_{0.2}Ti_{0.8}O_3$ [11] have been reported. Most studies of multiferroic heterostructures focus on FM/FE interface due to the limited ME effects away from it. However, our previous work demonstrates the thickness dependence of interfacial coupling in perovskite $SrTiO_3$ (STO)/$Pr_{0.67}Sr_{0.33}MnO_3$ (PSMO) heterostructures [12]. With the increase of PSMO thickness, the magnetic properties around the STO/PSMO interface changed due to the varying coupling of crystal structures with film thickness during the growth process. Considering the similarity between FM and FE materials of $3d$ transition-metal oxides, the structural and ferroelectric properties around the FM/FE interface may be modulated by FE film thickness, which could be exploited to study intrinsic ME effects in FM/FE heterostructures without an applied electric field.

Several mechanisms, including strain-mediated effects [13, 14], charge-driven screening effects [15, 16] and the exchange bias coupling [17, 18], have been proposed to account for the ME coupling in FM/FE oxide heterostructures. As reported previously [19], with increasing FE thickness the direction of self-polarization may be manipulated by flexoelectric effects through strain relaxation, causing the change of interfacial charge screening in the FM layer. At the same time, the strain effects accompanied with increasing FE thickness, also influence with the FM layer and modulate its crystal structure and electronic structure through the lattice mismatch, octahedral connectivity, atomic displacement, and orbital hybridization across the FM/FE interface [12, 20, 21]. The competition/cooperation of the charge-mediated and the strain-mediated effects stands out, yielding a robust ME response through the change of the lattice constants and the chemical valences [20, 22]. However, the gross change of the lattice constants sometimes cannot adequately reveal the important atomic information such as octahedral rotation [23] and preferential orbital occupancy [24], which control the electronic and magnetic properties of complex magnetic oxides. In order to have a better understanding of this entangled interaction, an in-depth investigation on the ME coupling mechanisms in FM/FE heterostructures is greatly demanded. Therefore, changing the FE thickness, which may cause the varying coupling of the charge screening at the FM/FE interface and the strain-mediated effects on local structures of FM layer, provides a viable means to study the ME effects in FM/FE thin film heterostructures.

In this work, we report an integrated study on the intrinsic ME coupling in (001) $SrTiO_3$ (STO)/$La_{0.67}Sr_{0.33}MnO_3$ (LSMO, fixed at ~14 unit cells) /$PbZr_{0.52}Ti_{0.48}O_3$ (PZT, 0-130 nm) heterostructures without an applied electric field. $La_{1-x}Sr_xMnO_3$, as a metallic bottom electrode, can be used to induce the self-poling of $PbZr_{1-x}Ti_xO_3$ thin films [25], which point downward to the substrate in this study. By increasing PZT thickness, the switching of self-polarization direction is realized through the competition between the interfacial effects and the strain relaxation-induced flexoelectric fields, resulting in noticeable effects on ME coupling in LSMO/PZT heterostructures. Thin PZT films (< 48 nm) drive the holes depletion at LSMO/PZT interface and preferential Mn 3$d$ out-of-plane orbital occupancy, enhancing magnetism of LSMO film. The strain-induced modifications of bond length/angle of $MnO_6$ associated with a modified magnetic spin configuration decrease the Curie temperature and magnetization of 14 unit cells LSMO film with thick PZT films (> 48 nm).

## II. EXPERIMENTAL

LSMO/PZT bilayer heterostructures were fabricated by pulsed laser deposition (PLD) on (001) SrTiO$_3$ (STO) single crystal substrates. LSMO thin film was deposited at 750°C under background oxygen pressure of 200 mTorr. The PZT thin film was grown on the top of LSMO at 550 °C under 50 mTorr oxygen pressure. The laser fluence was kept at 1~2 J/cm$^2$ and the pulse repetition frequency was 3Hz. After deposition, the sample was cooled down to room temperature at 15°C/min in a 1 Torr oxygen atmosphere. The thickness of LSMO was fixed at ~14 unit cells (u.c.), which was determined by considering both the dead layer thickness and interfacial charge screening length. It is known that metallic La$_{1-x}$Sr$_x$MnO$_3$ has a carrier density of 10$^{21}$-10$^{22}$/cm$^3$, yielding a screening length of less than 1 nm [26]. The presence of magnetically dead layer is well known for manganite perovskite and the thickness is found to be around 2 to 5 nm depending on the substrates [27]. Thus, LSMO of ~14 u.c. was chosen to reduce the impacts of dead layer and enhance the interface-to-volume ratio. PbZr$_{0.52}$Ti$_{0.48}$O$_3$ (PZT) is a good ferroelectric candidate in FM/FE heterostructures due to its high remnant polarization and outstanding piezoelectric coefficient [28]. The structural and ferroelectric properties of PZT depend on the film thickness, which was chosen as 4, 8, 16, 48, 96 and 130 nm.

Magnetic properties were measured by a superconducting quantum interference device (SQUID) and x-ray magnetic circular dichroism (XMCD), using beamline 4-ID-C of Advanced Photon Source (APS), Argonne National Laboratory, USA. Piezoelectric force microscopy (PFM) was used to measure the self-polarization of as-grown PZT thin films. Crystal structure was determined by high-resolution x-ray diffraction using synchrotron x-ray sources at the x-ray development and demonstration (XDD) beamline of Singapore Synchrotron

Light Source (SSLS) and beamline 12-ID-D of APS, Argonne National Laboratory, USA. The room temperature x-ray absorption near edge spectroscopy (XANES) measurement was performed using linear polarized x-ray at beamline 20-ID-B of the APS, Argonne National Laboratory, USA. The total fluorescent yield was detected and XANES normalization was conducted using the program Athena. High-angle annular dark field (HAADF) and annular bright field (ABF) in scanning transmission electron microscopy (STEM) and electron energy loss spectroscopy (EELS) were performed using the double aberration-corrected JEOL-ARM200CF microscope with a cold-field emission gun and operated at 200 kV, at Brookhaven National Laboratory, USA. The microscope is equipped with JEOL and Gatan HAADF detectors for incoherent HAADF (Z-contrast) imaging, Gatan GIF Quantum ER Energy Filter with dualEELS for EELS.

# III. RESULTS AND DISCUSSION

## A. Thickness-dependent self-polarization of PZT films

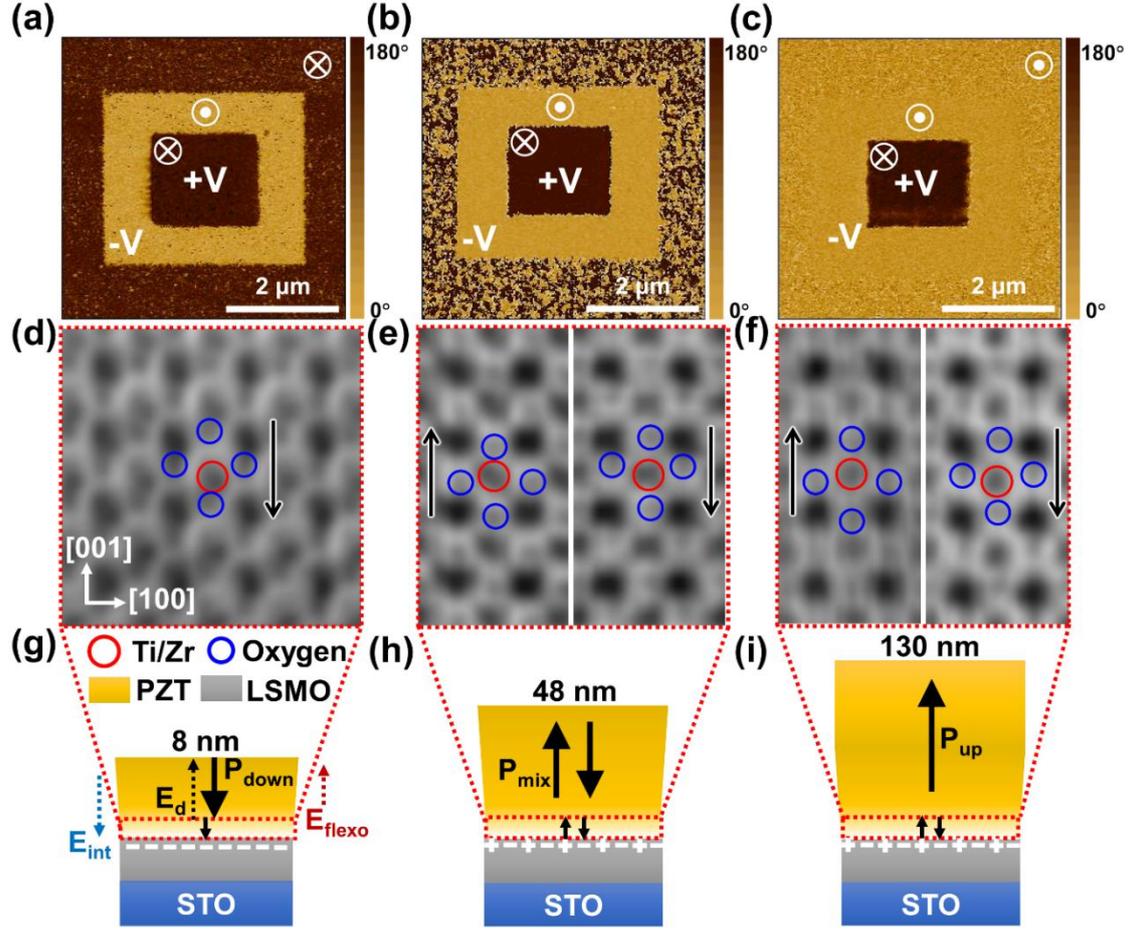

Figure 1. Room temperature out-of-plane PFM diagrams and STEM-ABF images viewed along [010] direction near the LSMO/PZT interface for samples with three different thicknesses of PZT: (a,d) 8 nm (b,e) 48 nm (c,f) 130 nm. $P_{down}$ is denoted by ⊗ and $P_{up}$ is denoted by ⊙. Panel (d)-(f) are the magnified portions of the STEM-ABF images of the PZT film near the LSMO/PZT interface (see the details in Figure S2). The arrows indicate the self-polarization direction near the LSMO/PZT interface, where LSMO in under the PZT layer. The red open circles correspond to Ti/Zr atoms and the blue open circles represent oxygen atoms. (g)-(i) Schematic illustrations of three sample configurations, showing the self-polarization induced charge screening in the interfacial LSMO. The plus sign indicates the hole and the minus sign corresponds to the electron. The large black arrows indicate the self-polarization direction of large thin film volume of PZT, while the small black arrows correspond to self-polarization direction near the LSMO/PZT interface marked by the red-dotted rectangle. $E_d$ is the depolarization field, $E_{int}$ is the built-in electric field and $E_{flexo}$ is the flexoelectric field.

As shown in Figures 1a-1c, the typical out-of-plane PFM poling and imaging [29, 30] were performed to probe the self-polarization of PZT as a function of its film thickness. The negative-voltage poled region corresponds to polarization pointing away from the substrate ($P_{up}$) while the positive-voltage poled region stands for polarization pointing down towards the substrate ($P_{down}$). The outermost part of diagram without an applied tip voltage represents the self-polarization state of as-grown PZT thin films. Therefore, the self-polarization of PZT gradually switches from $P_{down}$ to $P_{up}$ with increasing film thickness from 8 nm to 130 nm. For the 48 nm PZT, a mixed-polarization ($P_{mix}$) of $P_{down}$ and $P_{up}$ indicates the transition point of the switching of self-polarization. The detailed change of self-polarization as a function of PZT thickness and the experimental setup are shown in Figure S1. Since PFM results provide the ferroelectric properties in a large area, STEM was employed to study the local information of self-polarization. As shown in Figures 1d-1f, STEM-ABF images were viewed along the [010] direction from the PZT films with three different thicknesses, showing the projection of Ti/Zr (red open circles) and oxygen (blue open circles) atoms. The detailed information may be found in Figures S2a-S2c. At the portion of film near the LSMO/PZT interface, the self-polarization direction around the interface is clearly evident. It is found that $P_{down}$ and $P_{up}$ coexist in both 48 nm and 130 nm PZT, while the uniform $P_{down}$ state is presented in 8 nm PZT. These results are consistent with the out-of-plane PFM results, especially in the 8 nm and 48 nm PZT films. We notice that the probing depth in the PFM measurement dramatically depends on the dielectric permittivity of FE materials and contact conditions between the tip and sample surface [31, 32]. According to Kalinin et al [33], the tip-induced electroelastic field drops within a distance of approximately 100 nm for $PbZr_{1-x}Ti_xO_3$, resulting in a finite probing depth. Hence for 130 nm film, the PFM shows the upward self-polarization in the large thin film volume of PZT, which is not in conflict with the coexistence of $P_{down}$ and $P_{up}$ around the LSMO/PZT interface by STEM-ABF imaging.

Both PFM and STEM results reveal that the direction of self-polarization changes with PZT film thickness, as summarized in Figures 1g-1i. Several mechanisms, including the interfacial electric field induced by the bottom electrode, the evolution of the strain effects, the interface termination, and charged defects such as oxygen vacancies in FE films [30, 34-36], have been proposed for the FE self-polarization phenomenon. In the current study, the reversal of self-polarization as a function of PZT thickness may be interpreted as follows. It is well known that PZT thin film may suffer from the non-stoichiometric concentrations due to lead or oxygen vacancies during the deposition process [37, 38] and may be considered as a wide band-gap semiconductor [39]. The depletion region may form when the PZT contacts with the hole-doped metallic LSMO [40], which is analogous to the case of p-n/Schottky junction [41]. A simplified band alignment for the LSMO/PZT interface is illustrated in Figure S3, based on previous studies [42-44]. The depletion region would bend the band structure, forming the built-in electric field ($E_{int}$) directed from the PZT to the bottom electrode LSMO. Consequently, the $E_{int}$ may polarize the PZT into the $P_{down}$ state. Due to the low concentration of charge carriers in PZT and the high density of holes in LSMO film, the depletion region is mainly located in PZT and its width ranging from 3-35 nm depending on charge carriers densities, Ti/Zr ratio, and the ground state of the bottom electrode [39, 41]. This built-in electric field $E_{int}$ plays a dominate role in the self-poling of thin films PZT (< 48 nm), and there is an opposite depolarization field ($E_d$) accompanied with the self-polarization, as shown in Figure 1g. When the thickness of PZT increases, the interfacial effect gradually diminishes while the evolution of strain status has to be considered. We observe the strain relaxation in PZT with increasing film thickness, as shown in XRD results (see the details in Figures S4 and S5). The strain gradient points in the direction from the substrate to the film surface, thus generating a corresponding flexoelectric field ($E_{flexo}$) in the direction of the strain gradient, which favors the upward polarization [19]. This flexoelectric effect works in tandem with the $E_d$ against the interfacial effect, attempting to switch the self-polarization from the downward direction to the upward direction. Finally, for thick PZT films (> 48 nm), the interfacial effect is masked by the dominant strain gradient-induced flexoelectric field, driving the self-polarization into the $P_{up}$ state. The calculated changing trend of $E_{flexo}$ [45, 46] as a function of PZT thickness (Figure S6) supports the above discussion.

Combining above considerations, the changing trend of self-polarization in the whole thickness range is illustrated with a proposed competition between $E_{int}$ and $E_{flexo}$ in Figure S7. In the thinner PZT films, the large $E_{int}$ competes with a small $E_{flexo}$ and induces a downward self-polarization $P_{down}$ in 8 nm PZT. With increase of thickness, the strain relaxation occurs thus $E_{flexo}$ starts to increase and reaches the highest value at a critical thickness. The large $E_{flexo}$ would override the interfacial effect and turn the self-polarization orientation from pointing downwards to upwards. Therefore, a mixture of $P_{up}$ and $P_{down}$ is present in the 48 nm PZT and the region near the LSMO/PZT interface in the case of 130 nm PZT sample. With a further increase in film thickness, the $E_{flexo}$ begins to drop due to the totally-relaxed strain status in the thicker films and the self-polarization maintains upward, as shown in the large thin film volume of 130 nm PZT far away from the LSMO/PZT interface. Other possible effects, such as the depolarization field effect [34], charge on the PZT/air surface [44], and non-switchable layer of PZT [47] may need to be considered together with the effects of $E_{in}$ and $E_{flexo}$. These cooperative effects warrant further investigation and that is beyond the current scope of our paper. To sum up, the thickness could effectively tune the self-polarization of PZT films, and its effect on ME coupling in LSMO/PZT heterostructure is discussed in the following.

### B. Modulation of LSMO properties by PZT thickness

$La_{0.67}Sr_{0.33}MnO_3$ (LSMO) is chosen as a ferromagnetic component in the multiferroic heterostructure due to its good half-metallic behavior with a high curie temperature and a high spin polarization [48]. The PZT thickness-dependent self-polarization may electrostatically modulate the hole carrier density in LSMO and corresponding local structures across the interface, resulting in noticeable effects on LSMO properties. As shown in Figures 2a-2b, Curie temperature $T_c$ and the saturation magnetization $M_s$ vary coherently with increasing PZT thickness. $T_c$ was determined by finding the temperature for which the first derivative of the temperature-dependent magnetization curve has a minimum value (the inset of Figure 2a), while $M_s$ was acquired by the maximum magnetization under the 10 kOe magnetic field (Figure 2b). As summarized in Figure 2c, compared to the pure LSMO (0 nm PZT), the magnetic behaviors are first enhanced and then suppressed with increasing PZT thickness. Combined with above results of self-polarization, the $T_c$ and $M_s$ of LSMO are enhanced by 8 nm PZT in the $P_{down}$ state, whereas are weakened most by 130 nm PZT with the mixed self-polarization state at the LSMO/PZT interface. The variation of electrical properties of LSMO with increasing PZT thickness is in accordance with the changing trend of magnetism. The metal-insulator transition temperature $T_{MI}$, shown by the peak in the temperature-dependent resistivity curves, decrease by 25 K from the largest value (~ 309 K) in the $P_{down}$ state to the smallest one (~ 284 K) in the mixed state (Figure 2d). The detailed change of $T_{MI}$ as a function of PZT thickness is summarized in Table SI. In addition, the enhanced resistivity is clearly observed in the mixed state, as compared to that in the $P_{down}$ state. The electric field control of electrical properties is consistent with the PZT thickness-dependent $T_{MI}$ and resistivity of LSMO, as shown in Figures S8a-S8d. It is noted that the magnetic properties are linked to the Mn sites in LSMO, since only negligible contributions from Ti are found, as shown in Ti XMCD of Figure S9. The measurement of magnetic hysteresis loops at 120 K is shown in Figure S10. The variation of $M_s$ as a function of PZT thickness at 120 K is found to be consistent with that measured at 10 K.

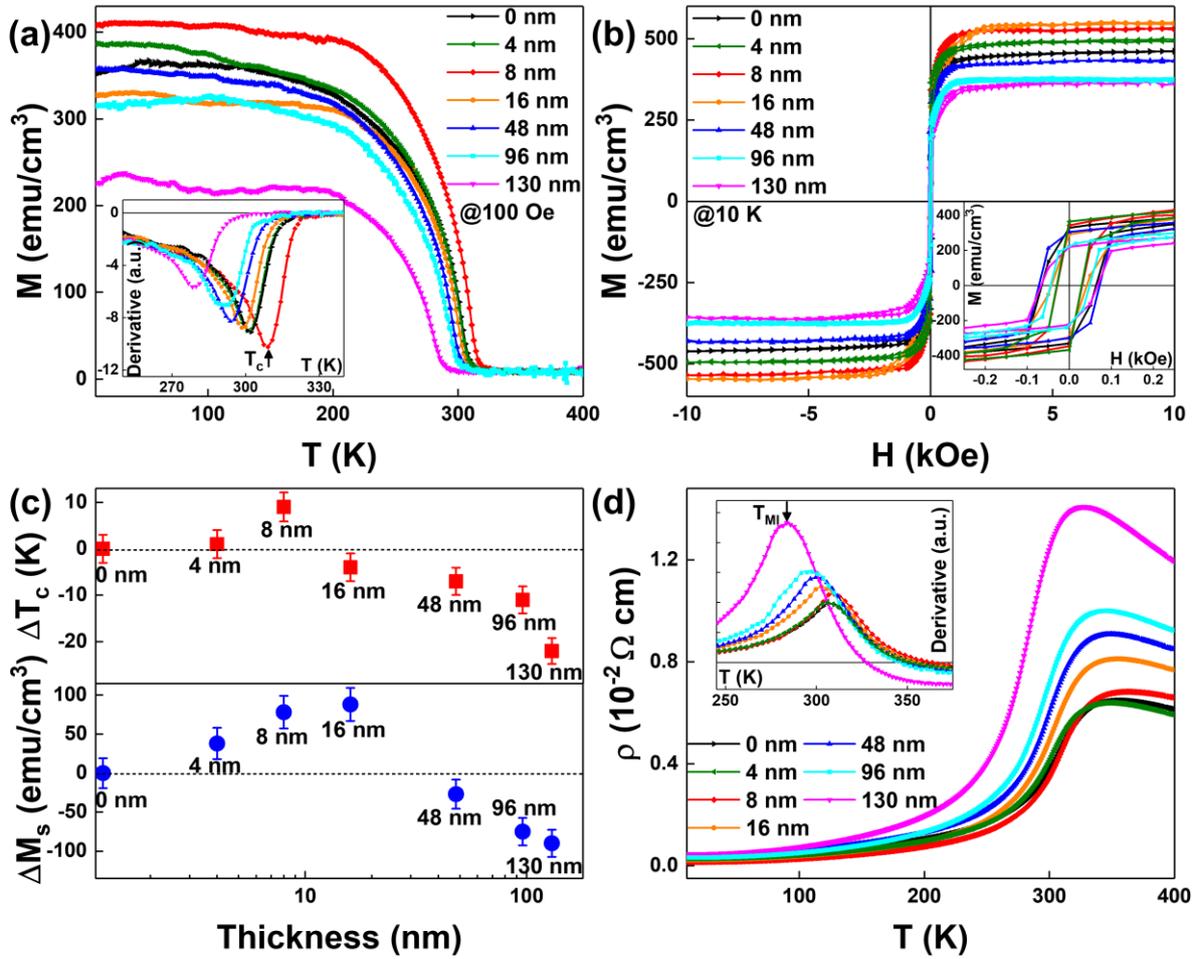

Figure 2. (a) Magnetization versus temperature (M-T) curves for LSMO with the different thicknesses of PZT with an applied 100 Oe magnetic field. The inset shows the first derivative of the M-T curve with a minimum value which is defined as Curie temperature $T_c$. (b) Magnetic hysteresis loops of LSMO with the different thicknesses of PZT measured at 10 K with an applied 10 kOe magnetic field. The inset shows the enlarged part of the magnetic hysteresis loops at low field ranges. (c) Summary of the $T_c$ difference ($\Delta T_c$) and the $M_s$ difference ($\Delta M_s$) between the pure LSMO (0 nm) and the LSMO with increasing thickness of PZT. (d) Resistivity versus temperature (R-T) curves for LSMO with the different thicknesses of PZT without an applied magnetic field. The inset shows the first derivative of the R-T curve with a maximum value which is defined as metal-insulator transition temperature $T_{MI}$.

The reversal of FE polarization-controlled magnetic and electrical properties of FM thin films has been reported. For temperature-dependent resistance cures of $La_{0.67}Sr_{0.33}MnO_3/BaTiO_3$ system by Cui et al [21], the enhanced resistivity and suppressed $T_{MI}$ of $La_{0.67}Sr_{0.33}MnO_3$ were observed when the overlying $BaTiO_3$ film in the $P_{up}$ state, as compared to the case of $P_{down}$. The enhanced resistivity is correlated with a modified spin exchange coupling, where an interfacial ferromagnetic metallic LSMO layer is antiferromagnetically coupled and therefore electrically decoupled from the bulk thin film layer. In our case, the interfacial charge carrier modulation driven by the self-polarization reversal and/or local structural distortions at the LSMO/PZT interface manipulated by overlying PZT films may give rise to this modified spin configuration. Meyer et al [25] reported the nearly 70% increase in the magnetization of $La_{0.8}Sr_{0.2}MnO_3/PbZr_{0.2}Ti_{0.8}O_3$ system when $La_{0.8}Sr_{0.2}MnO_3$ in holes depletion, and attributed this phenomenon to the direct control of charge carrier concentration by FE polarization. In our study, the reversal of self-polarization is realized by changing the PZT thickness, without the cumbersome design of devices required for application of an electric field. It is an attractive approach to observe intrinsic ME coupling in FM/FE heterostructures, and the origin of ME coupling in the current work is discussed as below.

### C. Interfacial charge-mediated effects

It has been demonstrated that the out-of-plane FE polarization is screened in the manganite within a few unit cells from the interface, leading to two possible interfacial states: depletion or accumulation of charge carriers [49]. Therefore, The PZT thickness-dependent self-polarization may electrostatically modulate the hole carrier density in LSMO, resulting in the depletion of holes at LSMO side of interface with 8 nm PZT and the mixture of electrons and holes at LSMO side of interface with 48 nm and 130 nm PZT, as illustrated in Figures 1g-1i. With a purpose of examining the interfacial-charge screening effect driven by self-polarized PZT films, we employed a room temperature XANES measurement on the entire thin film layer collected in total fluorescence yield (TFY) mode. Figure 3a shows the typical Mn $K$-edge XANES spectrum of LSMO thin film with different thicknesses of PZT films, consisting of a pre-edge peak (~ 6540 eV) and a sharp rising edge with a maximum resonance peak $E_r$ (~ 6555 eV). It is known that Mn $K$-edge main absorption edge arises from transition from the 1$s$ core level to unoccupied 4*p* orbitals, of which the edge position is found to be varied with the doping level *x* of mixed-valence manganite [50, 51]. In general, it shifts to the higher energy position when the $Mn^{4+}$ content increase. Compared with pure LSMO sample (0 nm PZT), a measurable shift of the inflection point to a lower energy position (~ 0.2 eV) is observed in LSMO with 8 nm PZT. Correlatively, an appreciable $E_r$ shifting to a lower energy (~ 0.3 eV) is recognized in LMSO with 8 nm PZT, together with an increased amplitude of this resonance peak. These observations indicate a decrease in the averaged valence state of Mn with 8 nm PZT, as compared to that of pure LSMO. It is in line with above results that 8 nm PZT in the $P_{down}$ state would electrostatically induces the accumulation of electrons (depletion of holes) in interfacial region of LSMO (Figure 1g). Based on the reported linear relationship [52] between the edge shift and the doping level *x*, $\Delta E = 3.0x$, an increase of 0.08 ± 0.03/Mn is estimated in averaged valence state for LSMO in holes depletion state compared to pure LSMO film. Since the magnetic moment changes by 1 $\mu_B$ when Mn changes from the low spin state $Mn^{4+}$ (S=3/2) to the high spin state $Mn^{3+}$ (S=2), the averaged increase of 0.08 ± 0.03 $\mu_B$/Mn is induced in LSMO in holes depletion state by valence modulation. Although the depletion interface (LSMO with 8 nm PZT) indeed enhance magnetism as supported by SQUID, this calculated change of Mn moment from the charge screening effect is smaller than that measured by SQUID (an increase of 0.46 ± 0.12 $\mu_B$/Mn), as shown in Figures 2b-2c. In addition, the decrease of magnetization for LSMO with 130 nm PZT is not consistent with the overall change of chemical valence.

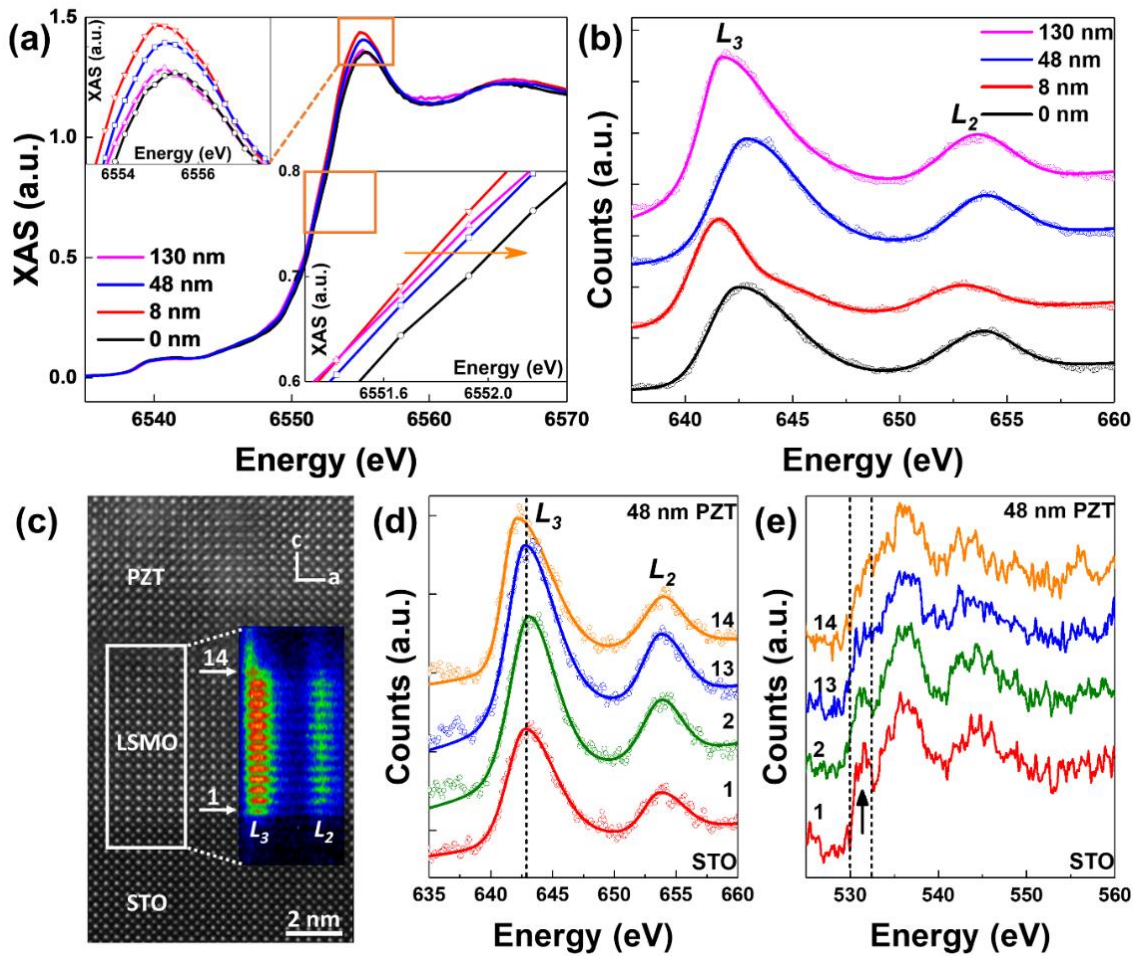

Figure 3. (a) The Mn $K$-edge XANES spectra of LSMO with 0 nm, 8 nm, 48 nm and 130 nm PZT, respectively. The insets show the enlarged part of the maximum peak and the position of inflection point. (b) EELS of Mn $L_{3,2}$-edge of LSMO with 0 nm, 8 nm, 48 nm and 130 nm PZT, respectively. (c) The STEM-HAADF image of STO/LSMO/PZT (48 nm) sample. The inset is the EELS spectral image of Mn $L_{3,2}$-edge averaged vertically from the area marked by the rectangle. The numbers indicate the position of atomic Mn columns starting from STO/LSMO interface. The number 1 corresponds to the first Mn column of LSMO at STO/LSMO interface and number 14 corresponds to the last Mn column of LSMO at LSMO/PZT interface. (d) Mn $L_{3,2}$-edge and (e) O $K$-edge EELS spectra of LSMO with 48 nm PZT, respectively, collected at selected Mn columns of LSMO near the STO/LSMO (number 1 and 2) and LSMO/PZT interface (number 13 and 14). The pre-peak feature is denoted by the arrow.

Then the detailed local information is obtained by Mn $L_{3,2}$-edge EELS. The chemical shift of Mn $L_{3,2}$-edge arises from the changes in the core-level energy and the effective charges on the atom [53]. As shown in Figure 3b, the Mn $L_{3,2}$-edge spectra (circles) were achieved by averaging the whole LSMO layer. The solid lines are the refinement by asymmetric Gaussian and Lorentzian function with refined $L_3$ peaks at 642.42 eV, 641.41 eV, 642.78 eV and 641.66 eV (from bottom to top) with an increase of PZT thickness, respectively. The standard error of each fitting was used to calculate the error bars (± 0.06 eV). Generally, the lower the $L_3$ energy position, the lower the averaged valence state of Mn [20, 22]. Therefore, LSMO with 8 nm PZT has the lowest valence state of Mn, which is well-matched with Mn *K*-edge XANES results. Compared with LSMO with 8 nm PZT, a noticeable positive shift for the 48 nm PZT sample (~ 1.37 eV) and a less significant one (~ 0.25 eV) for the 130 nm PZT sample are observed. Besides the averaged EELS information, the change of interfacial valence of LSMO was also investigated. Figure 3c shows the STEM-HAADF image of STO/LSMO (14 u.c.)/PZT (48 nm) sample and the atomic scale EELS mapping of Mn $L_{3,2}$-edge in the selected zone. The sharp interface may be observed with little Mn intermixing. These numbers indicate the position of atomic columns, where the Mn $L_{3,2}$-edge and the O *K*-edge fine structures were collected from. As shown in Figure 3d, in the case of LSMO with 48 nm PZT, the $L_3$ peak positions are 642.76 eV, 643.03 eV, 642.72 eV and 642.05 eV from the bulk LSMO (number 1,2) to the Mn columns near the LSMO/PZT interface (number 13,14). The error bars are ± 0.1 eV. The $L_3$ edge shifts to the lower energy position (~ 0.7 eV) from the bulk LSMO to the LSMO/PZT interface, indicating a lower valence state of interfacial LSMO than that of the bulk layer. This agrees with the proposed built-in electric field accompanied with the negative charge aggregation in the LSMO near the LSMO/PZT interface. Similar observations in LSMO with 8 nm and 130 nm PZT are shown in Figures S11a-S11b.

In addition, the O *K*-edge fine structures collected from number 1,2 and number 13, 14 are shown in Figure 3e. The pre-peak feature of the O *K*-edge, which is correlated with the hybridization of O 2*p* with Mn 3*d* bands, is sensitive to the oxidation state of Mn [54]. A linear relationship between the pre-peak intensity and the oxidation state of Mn in manganite oxides has been reported [55], where the pre-peak intensity increases when the nominal valence of Mn increases. Figure 3e shows that the pre-peak feature of LSMO with 48 nm PZT gradually diminishes from the bulk LSMO (number 1,2) to the LSMO/PZT interface (number 13, 14), also indicating a lower oxidation state of interfacial Mn than that of the bulk thin film layer. The suppression of pre-peak feature is also noticed in LSMO with 8 nm and 130 nm PZT near the LSMO/PZT interface, as shown in Figures S11c-S11d. In particular, the pre-peak feature is almost missing for whole LSMO thin film with 8 nm PZT (Figure S11c). In comparison, the visible pre-peak feature in the case of 48 nm PZT sample near LSMO/PZT interface (number 13 of Figure 3e) suggests a higher valence state of Mn than that of 8 nm PZT sample.

From the XANES, an averaged valence state of Mn is found to vary with PZT thickness-dependent self-polarization. LSMO with 8 nm PZT (in $P_{down}$ state) has the lowest valence state. Nevertheless, the enhanced magnetic properties from the interfacial charge screening may not be quantitatively explained by the valence-modulated spin state of Mn. In addition, the suppressed magnetic properties of LSMO with 130 nm PZT cannot be explained by the charge screening effect. The layer-resolved EELS results suggest the non-uniformity of the electronic structures in LSMO film due to the interfacial coupling. Considering the strong correlation between the electronic structure and crystal structure, additional mechanisms, such as the strain-mediated effects including the $MnO_6$ octahedral rotation and orbital reconstruction across the LSMO/PZT interface, may be responsible for the change of magnetism in our study.

### D. Strain-mediated effects

Due to the intrinsic strong coupling between the strain and polarization in FE materials, the strain-mediated effects may apply to LSMO through the lattice mismatch, octahedral connectivity, atomic displacement, and orbital hybridization across the LSMO/PZT interface [12, 20, 21]. According to the XRD results (Figures S4 and S5), all LSMO thin films are fully strained by the STO substrate. Considering the fixed thickness of LSMO, the strain effect from STO substrate is assumed to be constant for all the samples. In this study, we focus on the strain-mediated effects from overlying PZT thin films.

The depth profile of the out-of-plane lattice constants of LSMO, which vary across the LSMO/PZT interface, strongly depends on the thickness of overlying PZT films (see the details in Figure S12). The averaged out-of-plane lattice constants of LSMO are slightly changed (~0.5%) with increasing PZT thickness (Figure S4). The half-integer diffraction measurement, developed by Glazer in the 1970s [56, 57], was employed to characterize the $MnO_6$ octahedral rotation with increasing PZT thickness. As illustrated in Figure 4a, the presence of octahedral tilt results in the doubling of periodicity (2$d$) along certain axes [56, 58]. The half-integer peaks due to the oxygen scattering may be separated from the strong intensity of the cations scattering, since the periodic length of the cations remain as constant $d$. The use of high-flux synchrotron x-ray sources enables the accurate measurement of rotation pattern and the magnitude of octahedral tilts in perovskite thin films. Figure 4b shows a series of clear half-integer peaks in STO/LSMO/PZT (8 nm) heterostructure. The half-integer diffraction results of STO/LSMO, STO/LSMO/PZT (48 nm), and STO/LSMO/PZT (130 nm) heterostructures are shown in Figures S13a-S13c, respectively. By examining the presence/absence of the specific half-integer peaks, the octahedral rotation patterns of LSMO were identified as $a^-a^-c^-$ in all samples with PZT films according to the Glazer notation [57, 59].

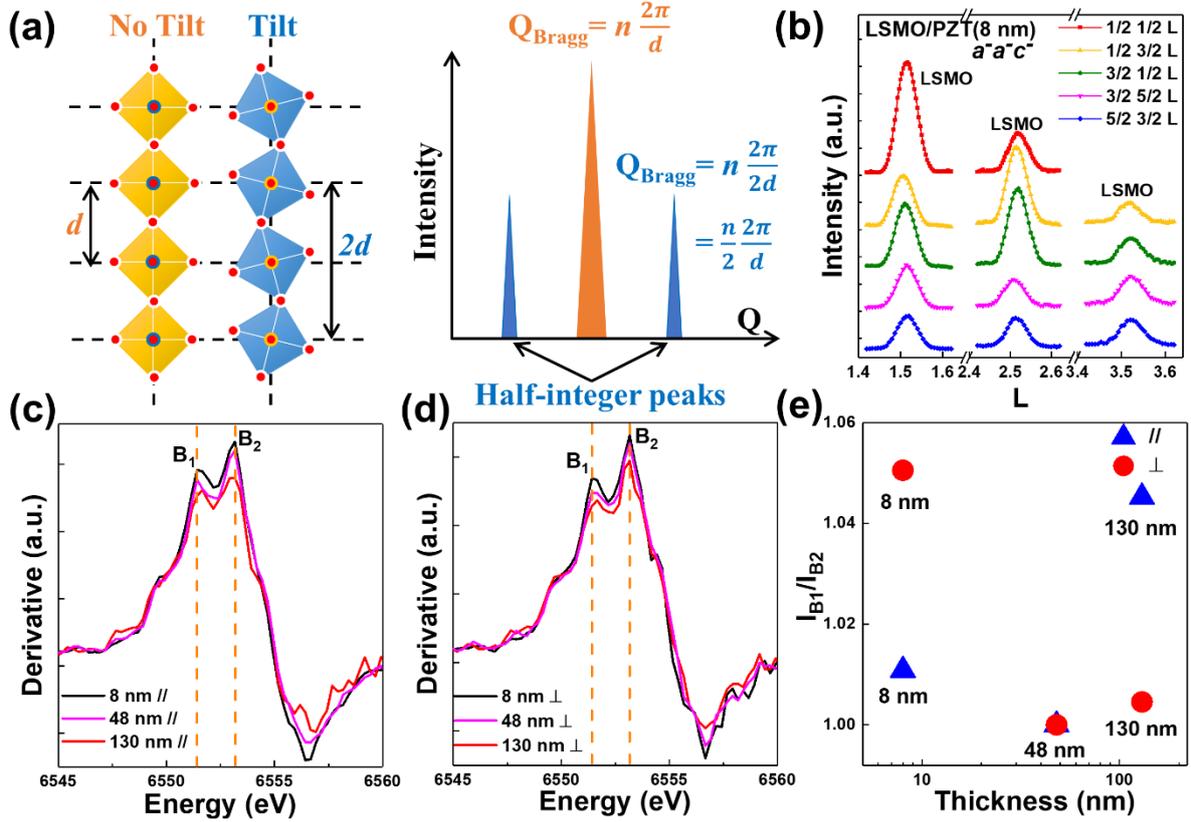

Figure 4. (a) Schematic illustration of the MnO$_6$ octahedral tilt and the appearance of half-integer peaks. Q is the transfer momentum in Bragg's law, *d* is the periodic length and *n* is an integer number determined by the order of diffraction. (b) Room temperature half-integer diffraction measurement of STO/LSMO/PZT (8 nm) heterostructure. The x-ray energy is 20 keV. The derivative curves of polarized Mn *K*-edge XANES spectra of LSMO with 8 nm, 48 nm and 130 nm PZT in (d) parallel (//) and (e) perpendicular (⊥) measurement. In parallel measurement, the *E* vector of incident x-ray is along the film plane; in perpendicular case, the *E* vector is perpendicular to the film plane. (f) Summary of the intensity ratio of B$_1$/B$_2$ in two polarization direction of Mn *K*-edge XANES spectra, normalized by the value of LSMO with 48 nm PZT for a comparison.

Based on the previous studies [60-62], the experimental half-integer peak intensities were used to calculate the tilt angles and oxygen positions by the home-made codes (see the details in Table SII and Figure S14). The calculated tilt angles, bond lengths and bond angles of MnO$_6$ with the error bars are summarized in the Table I. With an increase of PZT thickness, both the in-plane and the out-of-plane tilt angles of MnO$_6$ gradually decrease. The thicker PZT film seems to exert a damping effect on MnO$_6$ rotation, which induces the smallest out-of-plane bond length in LSMO with 130 nm PZT. Consistently, the ratio of the out-of-plane bond length to the in-plane bond length ($d_{out}/d_{in}$) is found to be smaller in LSMO with 130 nm PZT than that of LSMO with 8 nm and 48 nm PZT. In comparison, LSMO with 8 nm PZT has the largest out-of-plane bond length and $d_{out}/d_{in}$ ratio. The half-integer results reveal that anisotropic bond length of MnO$_6$ is induced by the PZT thickness-dependent strain-mediated effects, which may further affect the electron hopping and magnetic properties of LSMO.

Table I. The summary of the bond length ($d_{Mn-O}$), the bond angle ($\alpha_B$, $\gamma_B$) and the octahedral rotation angle ($\alpha_T$, $\gamma_T$) of LSMO with 8 nm, 48 nm and 130 nm PZT, along the in-plane and the out-of-plane direction. $d_{out}/d_{in}$ is the ratio of the out-of-plane $d_{Mn-O}$ to the in-plane $d_{Mn-O}$ value. The experimentally measured Curie temperature T$_c$ and the calculated electron hopping probability $t_{d-d}$ between neighboring Mn sites of LSMO with 8 nm, 48 nm and 130 nm PZT, normalized by the value of LSMO with 48 nm PZT.

| Sample | In-plane | | | Out-of-plane | | | $\dfrac{d_{out}}{d_{in}}$ | $\dfrac{T_c}{T_{c,48\,nm}}$ Exp. | $\dfrac{t_{d-d}}{t_{d-d,48\,nm}}$ Cal. |
|---|---|---|---|---|---|---|---|---|---|
| | $d_{Mn-O}$ (Å) | $\alpha_B$ (°) | $\alpha_T$ (°) | $d_{Mn-O}$ (Å) | $\gamma_B$ (°) | $\gamma_T$ (°) | | | |
| LSMO 8nm PZT | 1.972 (± 0.012) | 164.0 (± 2.8) | 5.8 (± 0.4) | 1.921 (± 0.011) | 163.9 (± 3.7) | 5.7 (± 0.9) | 0.974 (± 0.001) | 1.05 (± 0.01) | 1.03 (± 0.05) |
| LSMO 48nm PZT | 1.969 (± 0.012) | 165.1 (± 2.3) | 5.7 (± 0.7) | 1.919 (± 0.012) | 163.8 (± 3.4) | 5.0 (± 1.3) | 0.974 (± 0.001) | 1 | 1 |
| LSMO 130nm PZT | 1.967 (± 0.009) | 166.2 (± 3.2) | 5.0 (± 0.4) | 1.906 (± 0.008) | 165.9 (± 3.2) | 4.8 (± 0.5) | 0.969 (± 0.001) | 0.95 (± 0.01) | 0.97 (± 0.04) |

The polarization-dependent Mn $K$-edge XANES was then utilized to probe the strain-mediated effects on the electronic structure of LSMO, due to the fact that the detailed shape of curves is associated with the strong hybridization of Mn 3$d$ orbitals and O 2$p$ orbitals [63-65]. Figures 4c-4d show that, in two polarization directions, two sub-peaks (as denoted by B$_1$ and B$_2$) can be observed in the derivative curves of Mn $K$-edge XANES of all three samples. The lower energy peak B$_1$ corresponds to the 3$d^{n+1}\underline{L}4p^1$ electron configuration due to the Coulomb interaction between the Mn 1$s$ core hole (due to the x-ray photon excitations) and extra 3$d$ electron, whereas the higher energy peak B$_2$ corresponds to 3$d^n\underline{L}4p^1$ electronic state [61, 66]. Here, the *L* indicates the O ligands, and *L̲* means one hole in O ligands resulting from electron transfer from O 2*p* orbitals to Mn 3*d* orbitals. It is known that intensity ratio of B$_1$/B$_2$ is correlated to the strain-induced anisotropy of local structure environment, which may further make the hybridization between O 2*p* orbitals and Mn 3*d* orbitals anisotropic inside the MnO$_6$ octahedra. In general, the higher B$_1$/B$_2$ intensity, the larger Mn-O charge transfer probability. As summarized in Figure 4e, LSMO with 8 nm PZT has a higher B$_1$/B$_2$ value in the perpendicular measurement; in the parallel case, LSMO with 130 nm PZT has a higher B$_1$/B$_2$ value. LSMO with 48 nm PZT is taken as a reference with the lowest B$_1$/B$_2$ ratio in both polarization directions. Based on our previous work [61, 65], the dominant Mn-O charge transfer channel is the in-plane process (from O 2$p_x$, 2$p_y$ orbitals to Mn 3$d_{x^2-y^2}$ orbitals) for the perpendicular measurement and that is the out-of-plane process (from O 2$p_z$ orbitals to Mn 3$d_{3z^2-r^2}$ orbitals) for the parallel case. Therefore, LSMO with 8 nm PZT has a higher in-plane Mn-O hopping integral whereas LSMO with 130 nm PZT has a higher out-of-plane Mn-O hopping probability. The enhancement of charge transfer is in the direction where a larger overlap (i.e. smaller Mn-O bond length) between Mn 3*d* orbitals and O 2*p* orbitals. It is supported by half-integer results that d$_{out}$/d$_{in}$ is smaller in LSMO with 130 nm PZT than that of LSMO with 8 nm PZT.

The anisotropic Mn-O charge transfer is closely associated with preferential electron occupancy of Mn 3*d* orbital, which was taken into account for T$_c$ calculation [67]. According to the study of Chen et al [68], a larger Mn-O hopping probability results in a higher center-of-band energy and a less occupancy considering the fact that in LSMO two Mn *e*$_g$ states are antibonding in nature. In this context, LSMO with 8 nm PZT, with a larger *d$_{out}$/d$_{in}$* and higher in-plane Mn-O hopping, relatively prefers out-of-plane orbital occupancy; LSMO with 130 nm PZT, with a smaller *d$_{out}$/d$_{in}$* and higher out-of-plane Mn-O hopping, relatively prefers in-plane orbital states. The polarized XANES results provide the evidences of the strain-mediated effects on the Mn-O charge transfer and the preferential electron occupancy of Mn 3*d* orbitals, which is not easy to be detected by X-ray linear dichroism due to thick overlying PZT layer.

### E. Discussion

It is well-known that in the framework of double-exchange interaction, $T_c$ in metallic manganite is proportional to the electron hopping probability between two nearest Mn sites ($t_{d\text{-}d}$), which is determined by the orbital overlap between the Mn *3d* and O *2p* orbitals. The PZT thickness-dependent strain-mediated effects on the Mn-O bond length, Mn-O-Mn bond angle and Mn 3*d* orbital occupancy may modulate the electron hopping and therefore $T_c$ of LSMO, according to the following equation [61, 67]:

$$T_c \propto t_{d-d} = \sum_{i,j} \chi(\varphi_i)\chi(\varphi_j')V_{pd}^2$$

$$= \frac{1}{6}\{4\cdot\sin^2(\frac{\alpha_B}{2})\cdot d_{Mn-O,in}^{-7}\cdot n_{x^2-y^2}\cdot \chi(x^2-y^2)\cdot[\chi(x^2-y^2)+\chi(3z^2-r^2)]$$

$$+4\cdot\sin^2(\frac{\alpha_B}{2})\cdot d_{Mn-O,in}^{-7}\cdot(1-n_{x^2-y^2})\cdot \chi(3z^2-r^2)\cdot[\chi(x^2-y^2)+\chi(3z^2-r^2)]$$

$$+2\cdot\sin^2(\frac{\gamma_B}{2})\cdot d_{Mn-O,out}^{-7}\cdot(1-n_{x^2-y^2})\cdot \chi(3z^2-r^2)\cdot \chi(3z^2-r^2)\}  \quad (1)$$

where $\chi(\varphi_i)$ is the probability of the $e_g$ electron transferring from the $\varphi_i$ orbital of *i* Mn ion ($\varphi = 3d_{x^2-y^2}$ or $3d_{3z^2-r^2}$) to an O 2*p* orbital, and $\chi(\varphi_j')$ is the probability of the $e_g$ electron hopping from an O 2*p* orbital to the $\varphi_j'$ orbital in the neighboring *j* Mn ion. $V_{pd}$ is the matrix element between *p* and *d* orbitals and it is proportional to the $d_{Mn-O}^{-3.5}$ and $n_{x^2-y^2}$ is the ratio of electron occupancy in Mn $3d_{x^2-y^2}$ orbital. The transfer intensity $\chi(\varphi_i)$ and $\chi(\varphi_j')$ may refer to previous studies [67]. It is worth noting that $t_{d\text{-}d}$ is determined by not only the Mn-O bond length and Mn-O-Mn bond angle along both the in-plane and out-of-plane direction, but also the electron occupancy in Mn 3*d* orbitals as well. Based on the above discussion that LSMO with 8 nm PZT relatively prefers out-of-plane orbital occupancy while LSMO with 130 nm PZT relatively prefers in-plane orbital states, the assumptions of $n_{x^2-y^2} = 0$ for the LSMO with 8 nm PZT, 0.5 for the LSMO with 48 nm PZT, and 1 for the LSMO with 130 nm PZT, are used. Combined with the bond length/angle calculated from half-integer diffraction measurements, the calculated results of $t_{d\text{-}d}$ are shown in Table I. The changing trend of calculated values is well-matched with the experimental data, where the LSMO with 130 nm PZT has the smallest $t_{d\text{-}d}$ and LSMO with 8 nm PZT has the highest one. The small difference may result from the rough assumption of ratio of electron occupancy in Mn 3$d$ orbitals in three samples. This calculation demonstrates that PZT thickness-dependent strain-mediated effects on bond length/angle of MnO$_6$ and Mn 3$d$ orbital occupancy work in synergic to modulate T$_c$ of LSMO thin films.

As well known, the magnetic configuration depends on a delicate balance between the ferromagnetic and antiferromagnetic coupling. Additional effects, such as spin exchange coupling modified by the change of charge carrier density [69, 70], or interfacial orbital reconstruction [11, 21], could appear with the change of Curie temperature. The spin ordering in the strained manganese oxides is affected by the orbital preferential occupancy. According to the above results, the preferential Mn $3d_{x^2-y^2}$ ordering in LSMO with 130 nm PZT may stabilize the A-type antiferromagnetic (AFM) spin configuration, where spins couple ferromagnetically in the plane and antiferromagnetically along the out-of-plane direction [24]. It may lead to the depression of double-exchange mechanism and magnetization of LSMO [71]. In order to investigate the spin exchange coupling in LSMO/PZT heterostructures, measurements of zero field cooling (ZFC) and field cooling (FC) magnetization loops were conducted. Figures S15a-15c show no exchange bias in pure LSMO, LSMO with 8 nm and 48 nm PZT films, respectively. In comparison, a small exchange bias of ~1.7 Oe with an obvious enhancement of coercivity, as a typical phenomenon associated with the exchange anisotropy induced at the interface between AFM and FM layers, is observed in LSMO with 130 nm PZT (Figure S15d) [72]. Unlike the large exchange bias found in LSMO with antiferromagnetic $BiFeO_3$ films [73], the presence of antiferromagnetic spin coupling may arise from the LSMO/PZT interface through the orbital reconstruction. Therefore, the strain-mediated effects on bond length/angle of $MnO_6$ accompanied with a modified spin exchange coupling enables the PZT thickness-dependent modulation of magnetic properties of LSMO thin films. Apart from $T_c$, the decreased magnetization in LSMO with 130 nm PZT may be explained by the strain effects on the spin configuration of the Mn atoms.

By use of localized techniques such as STEM imaging and EELS (as summarized in Figure S16), we find the changes in interfacial lattice constants, chemical valences, and atomic displacements, which are combined to engender noticeable effects on properties of LSMO/PZT heterostructures. Polarized XANES and half-integer diffraction measurements are employed to probe the electronic anisotropy and modified bond lengths/angles of $MnO_6$, working in concert to illuminate the entangled interplay between the charge, spin, lattice and orbital in this strongly correlated system. Combining above strain- and charge-mediated effect, the self-polarization-driven carrier density modulation at the LSMO/PZT interface and the strain-mediated Mn $3d$ orbital occupancy work together to enhance magnetism of LSMO with PZT < 48 nm; the strain-induced the change of bond length/angle of $MnO_6$ accompanied with a modified spin configuration are responsible for the decrease in Curie temperature and magnetization of LSMO with PZT > 48nm.

## IV. SUMMARY

In summary, we develop the thickness-dependent self-polarization of PZT films, leading to noticeable effects on intrinsic magnetoelectric coupling in LSMO/PZT heterostructures. The competition model of the built-in electric field and the flexoelectric field is proposed to explain the self-polarization switching. With thin PZT films (< 48 nm), the self-polarization-driven valence modulation and strain-mediated Mn $3d$ orbital occupancy work together to enhance magnetism of LSMO thin films; with thick PZT films (> 48 nm), the strain-induced modifications of bond length/angle of $MnO_6$ associated with a modified spin configuration are responsible for the suppressed Curie temperature and magnetization in LSMO thin films. The study of competition/cooperation of charge-mediated and strain-mediated effects in LSMO/PZT heterostructures sheds light on the entangled interplay between the charge, lattice, orbital and spin in strongly correlated oxides. Our work provides a practicable means to induce desirable magnetoelectric effects in FM/FE heterostructures, with promising potentials in applications such as spintronic memory and logic devices.

# ACKNOWLEDGMENTS

The authors would like to acknowledge the Singapore Synchrotron Light Source (SSLS) for providing the facility necessary for conducting the research. The Laboratory is a National Research Infrastructure under the National Research Foundation Singapore. This research is supported by the Singapore Ministry of Education Academic Research Fud Tier 2 under the project No. MOE2015-T2-1-016 and the Singapore National Research Foundation under CRP Award No. NRF-CRP10-2012-02. Work at Brookhaven National Laboratory was supported by the U.S. Department of Energy, Office of Basic Energy Science, Division of Materials Science and Engineering, under Contact No. DESC0012704. Use of the Advanced Photon Source, an office of Science User Facility operated for the U.S. Department of Energy (DOE) Office of Science by Argonne National Laboratory, was supported by the U.S. DOE under Contact No. DE-AC02-06CH11357. We thank Dr. John W. Freeland for his kind support at beamline 4-ID-C of Advanced Photon Source.

Xiaoqian Yu and Lijun Wu contributed equally to this work.

# Supporting Information

## *For*

## Thickness-dependent self-polarization-induced intrinsic magnetoelectric effects in La$_{0.67}$Sr$_{0.33}$MnO$_3$/PbZr$_{0.52}$Ti$_{0.48}$O$_3$ heterostructures

Xiaoqian Yu[1], Lijun Wu[2], Bangmin Zhang[1], Hua Zhou[3], Yongqi Dong[3], Xiaohan Wu[1], Ronghui Kou[3], Ping Yang[4], Jingsheng Chen[1], Cheng-Jun Sun[3,*], Yimei Zhu[2,*], and Gan Moog Chow[1,*]

[1]*Department of Materials Science & Engineering, National University of Singapore, 117576, Singapore.*

[2]*Condensed Matter Physics & Materials Science Division, Brookhaven National Laboratory, New York 11973, USA.*

[3]*Advanced Photon Source, Argonne National Laboratory, Argonne, Illinois 60439, USA.*

[4]*Singapore Synchrotron Light Source, National University of Singapore, 117603, Singapore.*

*Corresponding Authors' Email: cjsun@aps.anl.gov; zhu@bnl.gov; msecgm@nus.edu.sg.*

# Outline

**S1. Out-of-plane PFM results with different thickness of PZT**

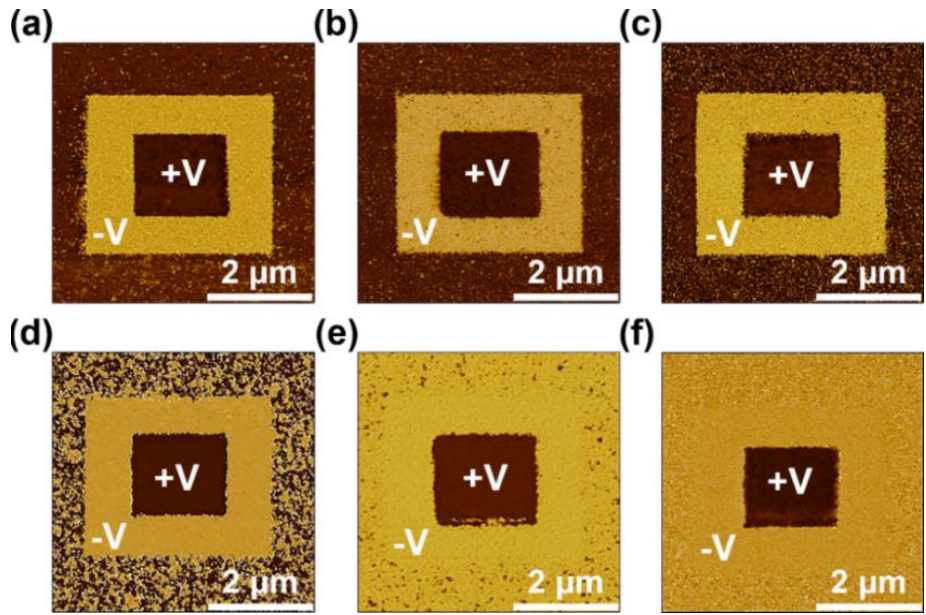

Figure S1: Room temperature out-of-plane PFM images of PZT with increasing film thickness: (a) 4 nm (b) 8 nm (c) 16 nm (d) 48 nm (e) 96 nm (f) 130 nm.

The PFM scan was first performed over a 3×3 µm² area with a negative-voltage tip bias, subsequently over a 1.5×1.5 µm² area with a positive-voltage tip bias at the center of the image, and finally over the 5×5 µm² area without a tip bias. As shown in Figures S1a-S1f, a distinct phase contrast is observed between the negative-voltage poled and the positive-voltage poled PZT thin film surface, manifesting a clear ferroelectric response in all PZT thin films. The color contrast represents the phase contrast, which is 180° switching in all the samples, demonstrating the out-of-plane polarization domain is oppositely switched between the negative-voltage poled region ($P_{up}$) and the positive-voltage poled region ($P_{down}$). The outermost part of the diagram without an applied tip voltage represents the spontaneous polarization state of as-grown PZT thin films. As shown in Figures S1a-S1f, the self-polarization of PZT gradually switches from $P_{down}$ to $P_{up}$ with an increase of thickness from 4 nm to 130 nm. Due to the different thickness of PZT film and the localized probe of a conductive tip that is in the contact mode, the coercive field and the applied voltage that is required to fully switch the domain are dependent on samples. The ± 5V, ± 3V and ± 8V was applied to 8 nm, 48 nm and 130 nm PZT, respectively.

## S2. STEM-ABF images near the LSMO/PZT interface

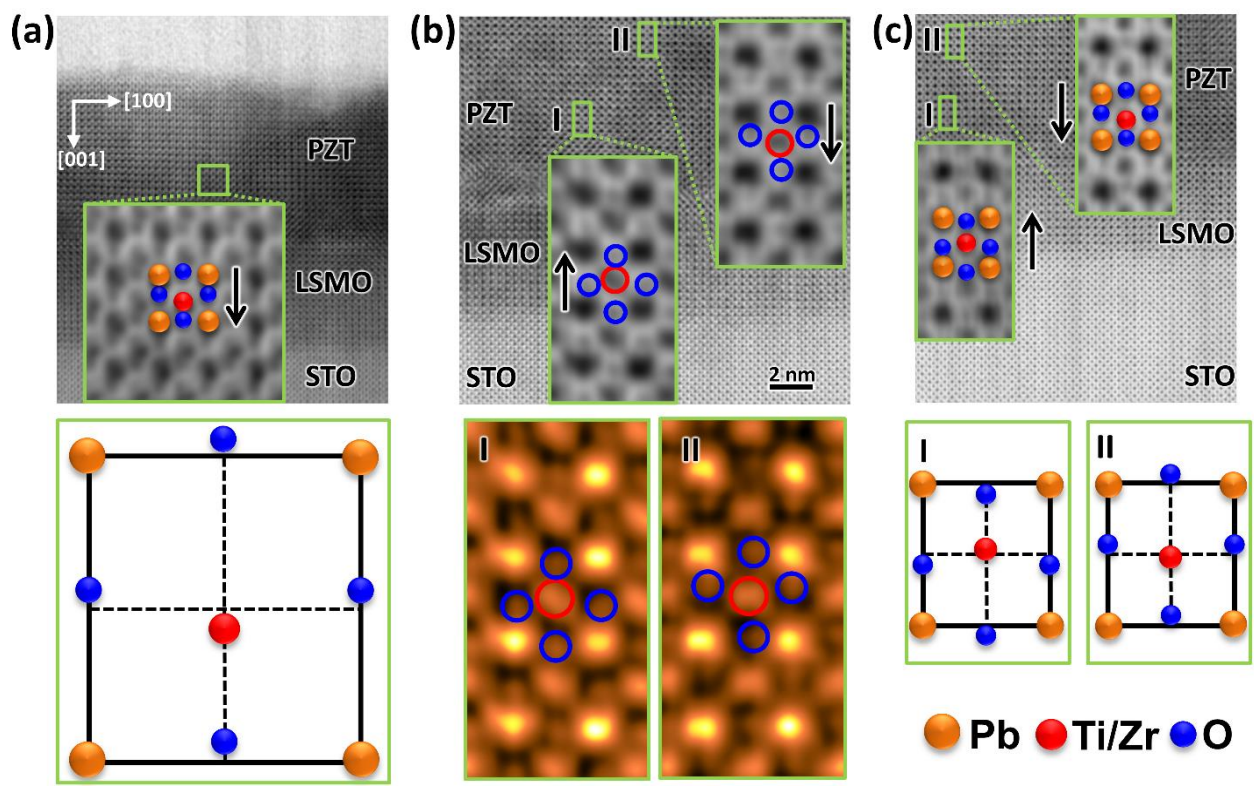

Figure S2: STEM-ABF images of (001) STO/LSMO/PZT heterostructures with different thickness of PZT: (a) 8 nm (b) 48 nm (c) 130 nm. The insets are the magnified portions of the STEM-ABF images of PZT film near the LSMO/PZT interface. The red open circles correspond to Ti/Zr atoms and blue open circles represent oxygen atoms. The schematic illustrations of 8 nm and 130 nm PZT atoms as filled circles are below (a) and (c), respectively. The simulation of STEM-ABF images of 48 nm PZT are below (b).

Figures S2a-S2c show the STEM-ABF images of (001) STO/LSMO/PZT with 8 nm, 48 nm and 130 nm PZT, respectively, indicating the self-polarization direction around the LSMO/PZT interface. The insets are the magnified STEM-ABF images of PZT in the selected zones, as shown in Figures 1d-1f in the main text. The darker black circles in the image are the STEM data representing the metal atoms, and the lighter black circles correspond to oxygen atoms. Below Figure S2a and Figure S2c, respectively, the schematic illustration of 8 nm and 130 nm PZT as filled circles are drawn to help guide the self-polarization

direction. In Figure S2b, the open circles (red and blue) are drawn in order to help better visualize the atom positions. Below Figure S2b, the simulation of STEM-ABF images of 48 nm PZT show the coexistence of $P_{up}$ and $P_{down}$ state. The displacement of atoms may also be observed throughout the entire STEM-ABF image under consideration. The weak signal for oxygen is due to its small atomic number and the resolution limit in our current STEM imaging. We used the home-made computer software to filter the raw data in order to get a higher-quality signal for analysis.

## S3. Proposed band diagram at the LSMO/PZT interface

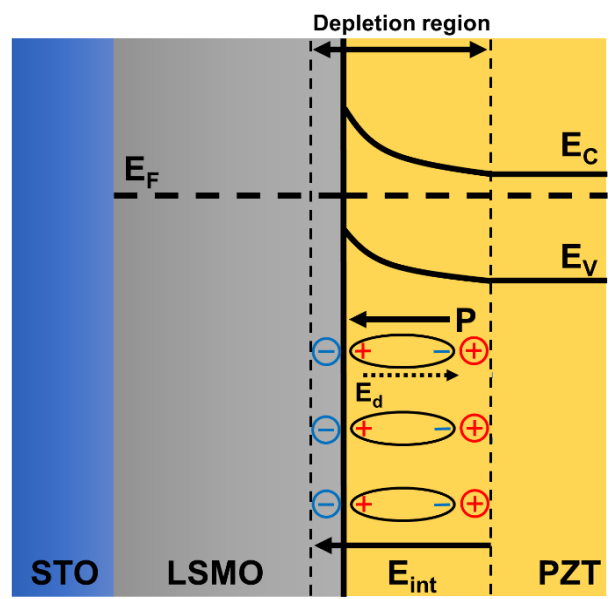

Figure S3: Proposed band diagram for the LSMO/PZT interface. $E_C$, $E_V$, and $E_F$ indicate the conduction band, valence band and Fermi level, respectively.

As shown in Figure S3, a simplified band diagram was proposed for the LSMO/PZT interface. The electrostatic equilibrium requires the same Fermi level of LSMO and PZT. Since the target we used contains a 20 mol% excess of PbO and growth oxygen pressure is 50 mtorr, the excess Pb may extract mobile oxygen from the perovskite matrix in oxidation and therefore the formation of oxygen vacancies may exist as charge carriers in PZT films [1, 2]. Here, we assume that PZT is a n-type semiconductor in this simplified band diagram. When PZT contacts with the metallic bottom electrode LSMO (hole-doped), as analogous to the p-n/Schottky junction, the depletion effect would bend band structure, leading to the positive charge diffusion into PZT and the negative charge aggregation in the interfacial LSMO and therefore forming a built-in electric field ($E_{int}$) directed from the PZT to the bottom electrode LSMO. Consequently, this $E_{int}$ of the depletion region may polarize the PZT into $P_{down}$ state. As illustrated in Figure S3, in the case of $P_{down}$, there will be an opposite depolarization field ($E_d$) accompanying with the downward polarized domains. The schematic illustrations of $E_{int}$ and $E_d$ in LSMO with 8 nm, 48 nm and 130 nm PZT are shown in Figure S7. Other factors such as the charge on the PZT/air surface and other charged defects were not considered in this simplified band diagram, which requires further study and that is beyond the current scope of this paper.

## S4. XRD *θ-2θ* scans results

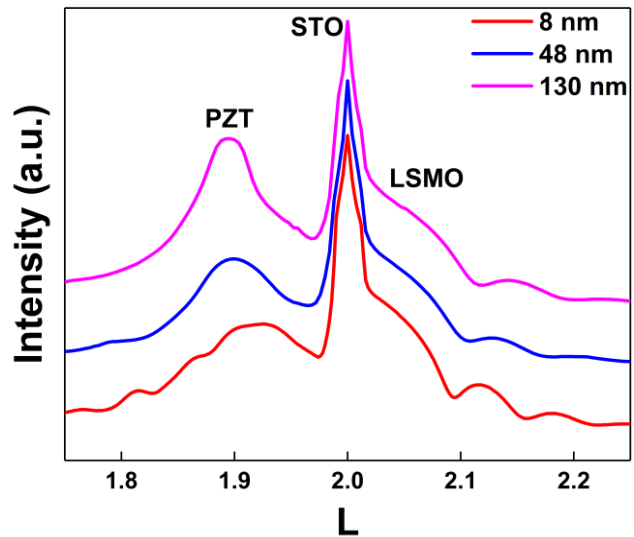

Figure S4: XRD *θ-2θ* scans for STO/LSMO (~14 u.c.)/PZT (8 nm, 48 nm, and 130 nm) heterostructures.

With increasing PZT thickness, both LSMO and PZT remain highly *c*-axis orientated, which indicates a good epitaxial growth of thin film layer on (001) STO substrate. Even the difference of out-of-plane lattice constants between LSMO film and STO substrate ($c_{STO}$ = 3.905 Å) is not small, the small thickness of LSMO makes it hard to separate the film and the substrate peak. The LSMO peak position may be determined by thickness oscillation. Out-of-plane lattice constants of LSMO with 8 nm, 48 nm and 130 nm PZT were calculated as 3.804 Å, 3.799 Å and 3.784 Å, respectively. An estimation shows a slight difference (~0.5%) in the averaged out-of-plane lattice constants of LSMO, induced by the lattice mismatch from topping PZT layers with increasing thickness.

The out-of-plane lattice constants of PZT are derived from the (002) diffraction angle as 4.124 Å, 4.111 Å and 4.076 Å for 130nm, 48nm and 8nm samples, respectively. The out-of-plane lattice constant of 130 nm PZT film is close to the value of bulk PZT ($c_{tetragonal}$ = 4.139 Å) [3], whereas that of 48 nm and 8 nm PZT are unexpectedly small. Considering that the Ti/Zr ratio of PZT in our system is at the morphotropic phase boundary (MPB) [4], the introduction of a secondary phase in PZT is suggested with the support of appreciable humps residing at the right shoulder of the PZT peaks (especially in 8 nm sample). According to computed variation of crystal lattice parameters as a function of PZT composition ratio, the lattice constant linearly decreases in the rhombohedral region and the lattice parameter is found as $a_r \approx 4.075$ Å near the MPB of bulk PZT [3, 5]. This value is close to the out-of-plane lattice constant of 8 nm PZT. It was experimentally demonstrated that the thickness dependent transition from a tetragonal structure to a rhombohedral structure in $PbZr_{0.52}Ti_{0.48}O_3$ [6], even in $PbZr_{0.35}Ti_{0.65}O_3$ of which the bulk form of this composition ratio should be tetragonal [7]. The relative weight of rhombohedral and tetragonal structure was found to vary with film thickness, where thinner PZT were more rhombohedral and the tetragonal structure dominated in thicker film even with the rhombohedral PZT at the bottom interface. In our current work, the thickness-dependent weighted rhombohedral structure is responsible for the unexpectedly small lattice constants of 8 nm and 48 nm PZT and the tetragonal structure predominated in 130 nm PZT with bulk-like lattice parameters (as supported by reciprocal space mappings results in Figure S5).

## S5. XRD reciprocal space mapping results

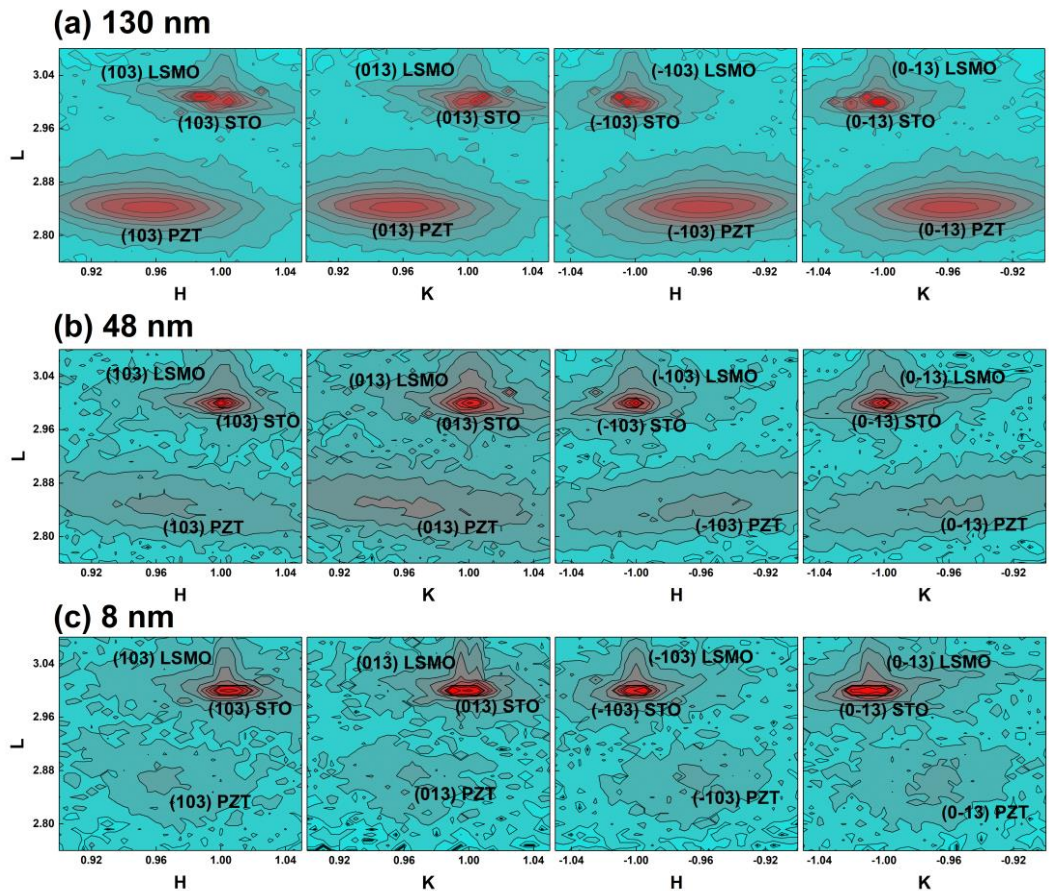

Figure S5: Reciprocal space mappings (RSM) around of STO/LSMO/PZT with different thickness of PZT: (a) 130 nm (b) 48 nm (c) 8 nm.

In order to investigate the strain status and in-plane lattice constants of LSMO and PZT films on STO substrate, reciprocal space mappings using high-resolution XRD were conducted (Figures S5a-S5c). All the LSMO thin films are completely strained on the STO substrate, inferred from the vertical alignment of reciprocal lattice points of LSMO and STO. Thus, the in-plane lattice constants of LSMO are the same as those of substrate. The tetragonal structure is determined by the same $L$ values in all {103} mappings of LSMO in three samples.

The vertical displacement of PZT reciprocal lattice points with those of STO demonstrates that all PZT thin films are relaxed. This is possibly due to the misfit dislocations at interface and oxygen vacancies in PZT films. The in-plane lattice constants are determined from RSMs as $a=b=4.078$ Å, 4.068 Å, 4.047 Å, for 130 nm, 48 nm and 8 nm PZT, respectively. The out-of-plane lattice constants determined from RSMs are 4.125 Å, 4.114 Å and 4.079 Å for 130 nm, 48 nm and 8 nm PZT, respectively, in consistent with the values calculated from XRD $\theta$-$2\theta$ results. According to the lattice parameters of bulk tetragonal $PbZr_{0.52}Ti_{0.48}O_3$ ($a_{tetragonal}$ = 4.046 Å, $c_{tetragonal}$ = 4.139 Å), the coexistence of tetragonal and rhombohedral phase is proposed in PZT films [3, 5]. The relative weight of rhombohedral and tetragonal structure is found to be varied with film thickness. The dominant tetragonal structure is indicated by the same $L$ values in all {103} mappings of 130 nm and 48 nm PZT films, whereas a larger weight of rhombohedral phase is suggested in 8 nm PZT. A small difference in $L$ values of (013) and (0-13) can be seen in 8 nm PZT, where $L=2.872$ for PZT (013) and $L=2.856$ for PZT (0-13), suggesting a non-90° angle between direct lattice constant $b$ and $c$. Similarly, a minor difference in $L$ values of PZT (103) and PZT (-103), where $L=2.864$ for PZT (103) and $L=2.872$ for PZT (-103), indicates a non-90° angle between direct lattice constant $a$ and $c$.

## S6. Estimation of strain gradients and $E_{flexo}$ in PZT

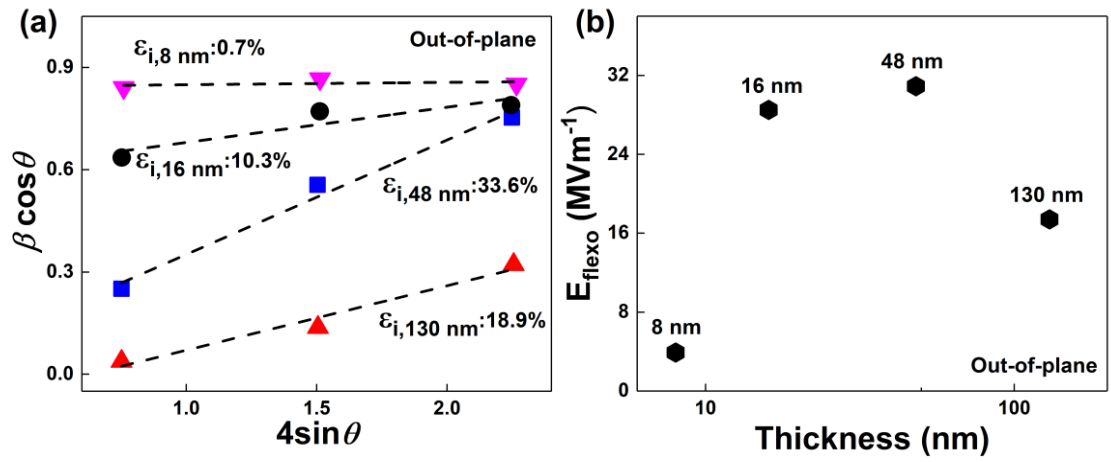

Figure S6: (a) Out-of-plane Williamson-Hall plots for the inhomogeneous strain ($\varepsilon_i$) of the PZT films; (b) Out-of-plane flexoelectric field as a function of PZT thickness.

The FE self-polarization direction depends on film thickness and strain relaxation. We proposed that the strain gradient-mediated flexoelectric effect may reverse the downward self-polarization direction of PZT to an upward self-polarization in thick films. From the above RSM results, the structural relaxation occurs in all PZT films. According to the quantitative analysis of relaxation [8], 8 nm PZT is partially relaxed while 130 nm PZT are totally relaxed. We assumed that strain penetration lengths in PZT films are 8 nm for 8 nm PZT, 16 nm for 16 nm PZT and 48 nm for 48 nm PZT and 130 nm PZT.

In order to obtain more quantitative information, we estimated the strain gradients by Williamson-Hall (W-H) plots [9]. The W-H plots of PZT thin film were obtained from XRD $\theta$-$2\theta$ peak widths. Three peaks (001), (002) and (003) from $\theta$-$2\theta$ scans were used to determine the out-of-plane change of strain status in PZT. According to previous studies [10], the strain gradient is derived from inhomogeneous strain, $\varepsilon_i$, which could be determined from the slope of the fitted equation (the dash line). As shown in Figure S6a, $\beta$ corresponds to the peak width, mainly contributed from the inhomogeneous strain $\varepsilon_i$ and finite thickness of samples. The $\varepsilon_i$ values are determined as 0.7%, 10.3%, 33.6% and 18.9% for 8 nm, 16 nm, 48 nm and 130 nm PZT, respectively. From the inhomogeneous strain values and the assumption of strain penetration length, we estimated the out-of-plane strain gradients as $-8.7 \times 10^5$ m$^{-1}$, $-6.4 \times 10^6$ m$^{-1}$, $-7.0 \times 10^6$ m$^{-1}$ and $-3.9 \times 10^6$ m$^{-1}$ for 8 nm, 16 nm, 48 nm and 130 nm PZT, respectively. Based on the reported equation [11], the flexoelectric field ($E_{flexo}$) values were determined as 3.9 MVm$^{-1}$, 28.5 MVm$^{-1}$, 30.9 MVm$^{-1}$ and 17.4 MVm$^{-1}$ for 8 nm, 16 nm, 48 nm and 130 nm PZT, respectively. Figure S6b shows the changing trend of $E_{flexo}$ as a function of PZT thickness, which is consistent with our proposed mechanism in main text. The calculated values are larger than the published data in previous reports [12]. We note that the presence of secondary phase and defects such as oxygen vacancies in PZT thin film may contribute to the peak broadening, thus increasing the $\varepsilon_i$ and strain gradient.

## S7. The competition between $E_{int}$ and $E_{flexo}$ as a function of PZT thickness

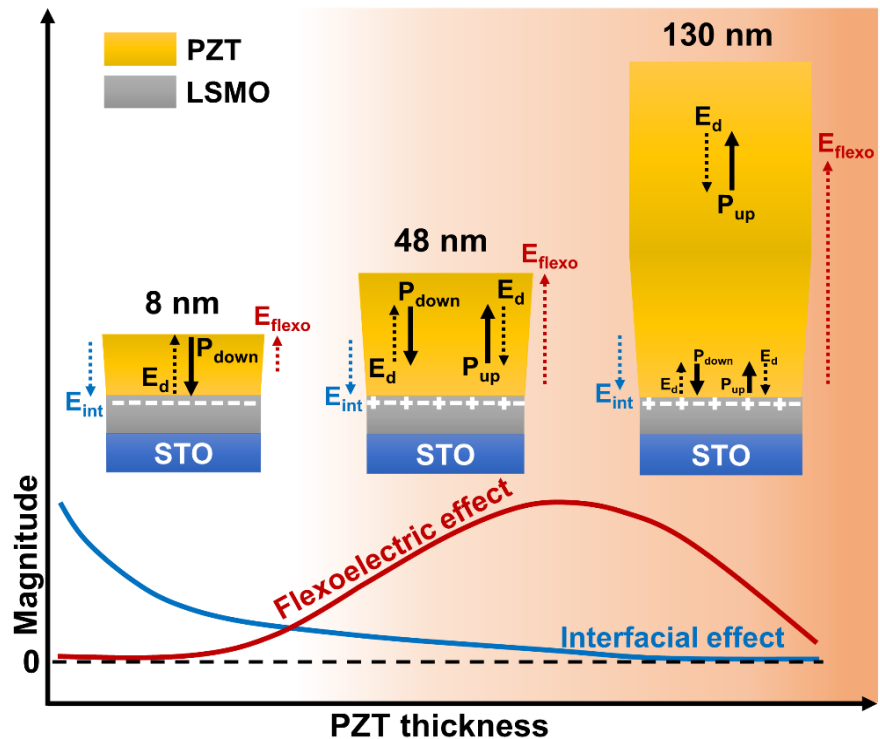

Figure S7: Schematic illustrations of three samples configurations and the competition between $E_{int}$ and $E_{flexo}$ as a function of PZT thickness. $E_d$ is the depolarization field, $E_{int}$ is the built-in electric field, and $E_{flexo}$ is the flexoelectric field. The solid arrows indicate the polarization direction and the dotted arrows correspond to the electric fields. The magnitude of the self-polarization and the electric fields are independent of the arrow size.

The magnitude of the interfacial effect of $E_{int}$ (blue curve) and the flexoelectric effect $E_{flexo}$ (red curve) as a function of PZT thickness are shown in Figure S7. The changing trend of $E_{flexo}$ was roughly estimated based on the Figure S6b. It is noted that the calculated $E_{flexo}$ from XRD results is the averaged-volume value and therefore the localized techniques such as the SETM may be required for a more accurate description of the strain gradient-driven flexoelectric effect in PZT film.

## S8. Electric field control of electrical properties of LSMO

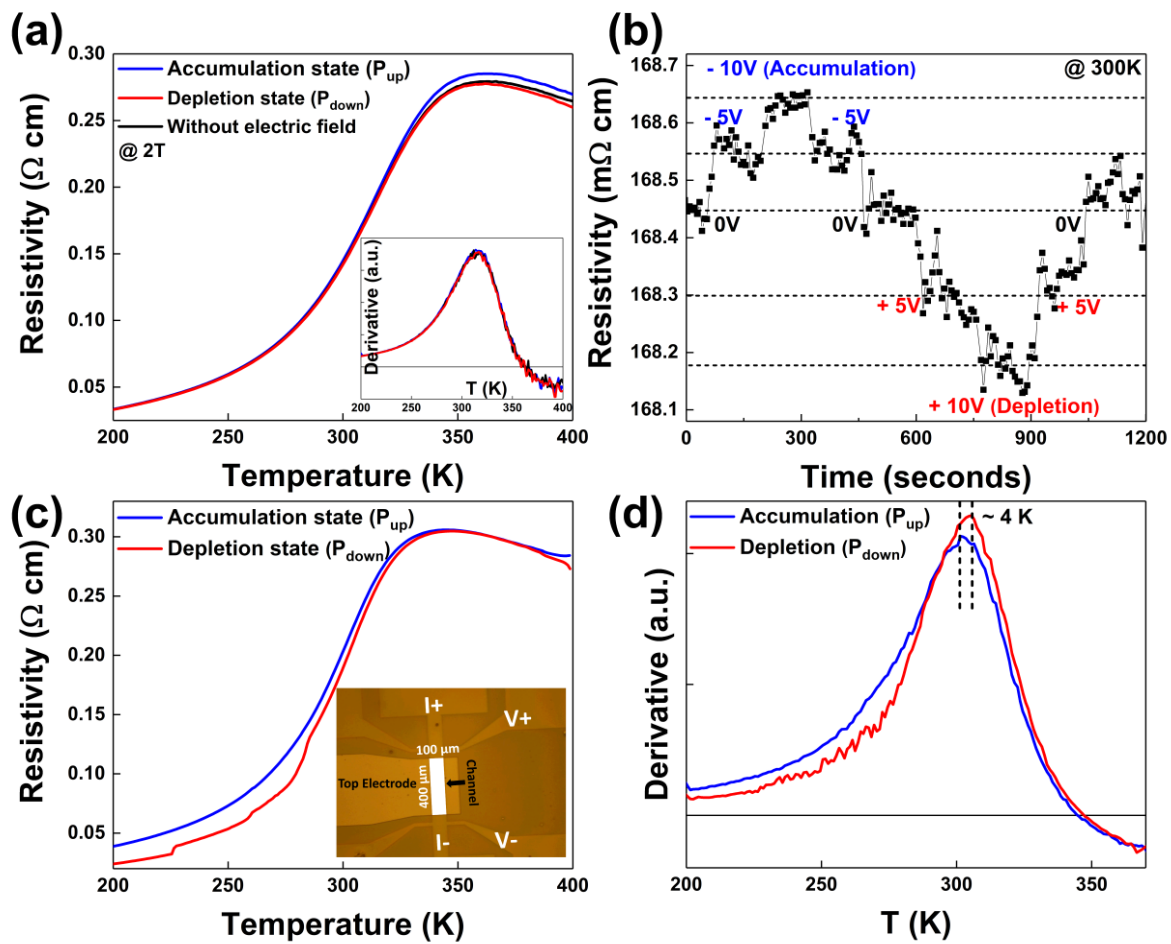

Figure S8: Electric field control of electrical properties of STO/LSMO (~ 30 u.c.)/PZT (~ 130 nm) thin film heterostructures. (a) Resistivity versus temperature (R-T) curves for the accumulation state (applied with -10 V), depletion state (applied with +10 V), and zero voltage state at 2 T magnetic field. The inset shows the first derivative of R-T curves. (b) Room temperature real-time measurement of resistivity with applied voltages. A positive voltage induces PZT into the $P_{down}$ state and a negative voltage induces PZT into the $P_{up}$ state. (c) R-T curves measured for the accumulation state and depletion state without an applied magnetic field. The inset shows the device structure indicating the channel area of Hall bar geometry. (d) The first derivative of R-T curves in (c), where the metal-insulator transition temperatures denoted by dash lines.

Figures S8a-S8b show that the resistivity is enhanced when LSMO in the accumulation state, as compared to that of the depletion state and zero voltage state. The metal-insulator transition temperature remains unchanged with applied voltages (as shown in the inset of Figure S8a). After making samples into the devices, the R-T curves were measured without an applied magnetic field (Figure

S8c). In Figure S8d, a slight increase (~ 4 K) of metal-insulator transition temperature is observed for LSMO in the depletion state. This is possibly due to fewer leakage currents generated (i.e. more voltages applied to the PZT thin film layer) in a device with a Hall bar structure. The electric field control electric properties of LSMO is consistent with our observation of PZT thickness-dependent electric properties of LSMO, where LSMO with 8 nm PZT (LSMO in the depletion state) has a smaller resistivity and a higher metal-insulator transition temperature than that of LSMO with thicker PZT films.

## S9. Ti *L*-edge XMCD

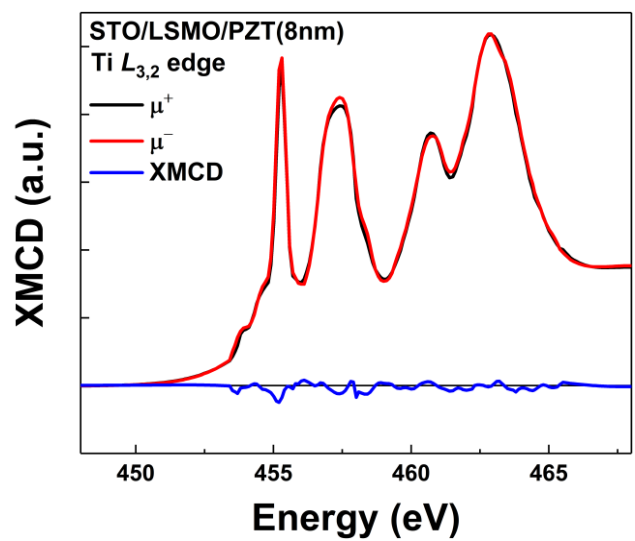

Figure S9: Ti *L*-edge XMCD for STO/LSMO/PZT (8 nm) under 5000 Oe magnetic field at 80K.

The absence of XMCD signal provides a clear evidence of nonmagnetic contribution from Ti atoms. Magnetic moments in this work arise from the Mn sites in LSMO. The signals were collected from bulk-sensitive total florescence yield (TFY) mode.

## S10. Magnetic hysteresis loops at 120 K

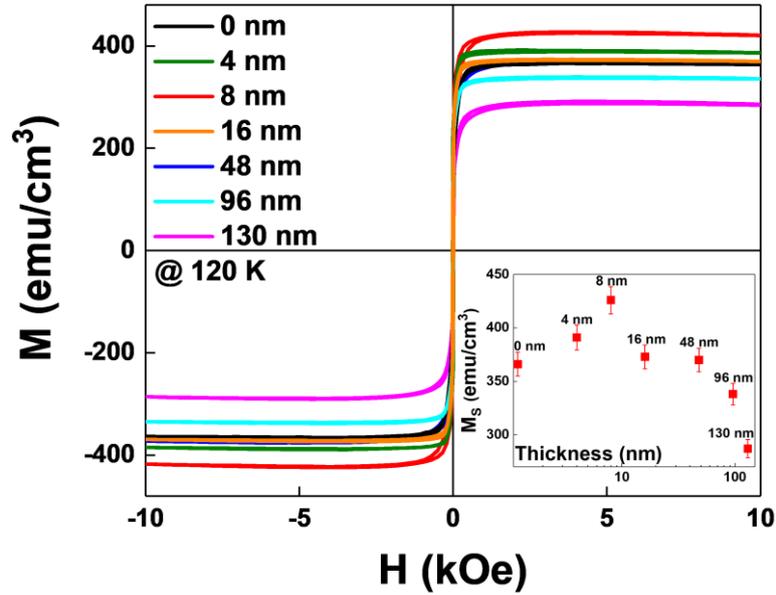

Figure S10: Magnetic hysteresis loops of LSMO under the different thickness of PZT measured at 120 K with an applied 10 kOe magnetic field. The inset shows the variation of $M_s$ with the error bars as a function of PZT thickness.

The saturation magnetization $M_s$, which is extracted from the magnetic hysteresis loops measured at 120 K, is first enhanced and then suppressed with an increase of PZT thickness. This changing trend is found to be consistent with that observed at 10 K (Figure 2c).

## S11. Mn $L_{3,2}$-edge and O $K$-edge EELS spectra

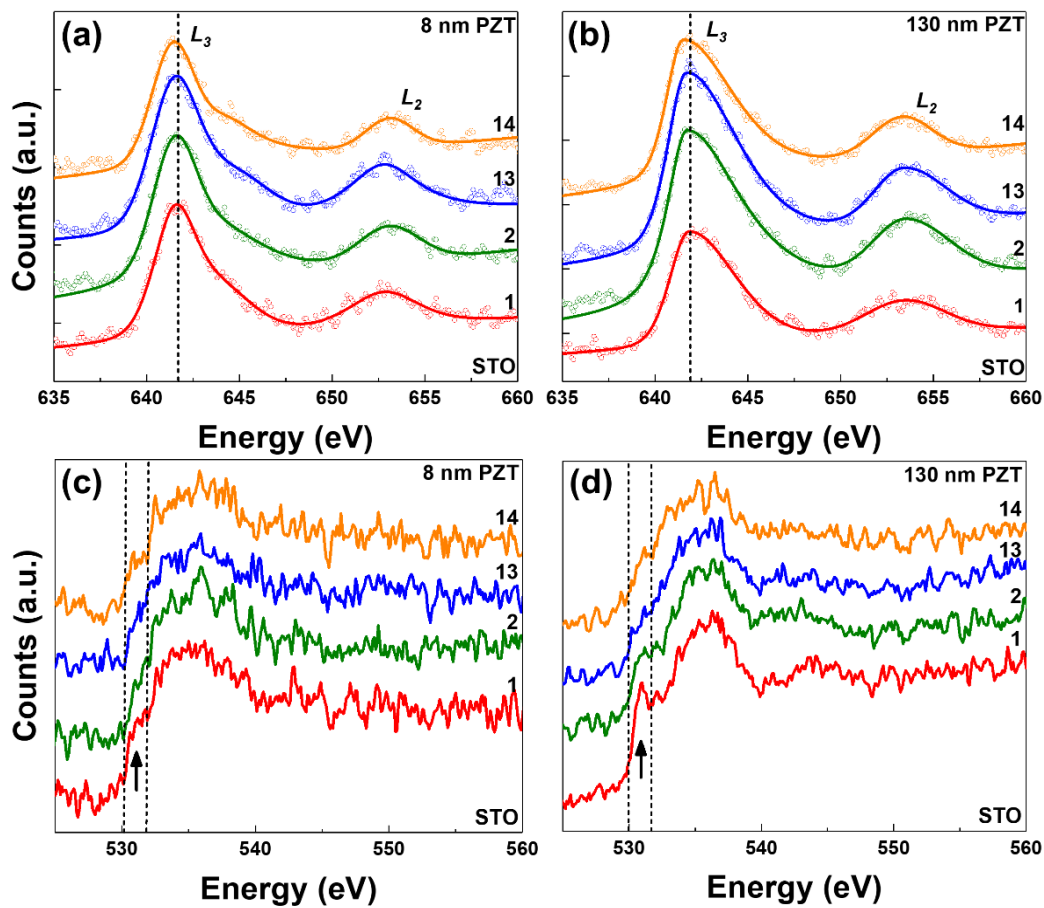

Figure S11: (a-b) Mn $L_{3,2}$-edge and (c-d) O $K$-edge EELS spectra of LSMO with 8 nm and 130 nm PZT, respectively, collected at selected Mn columns of LSMO near the STO/LSMO interface (number 1 and 2) and LSMO/PZT interface (number 13 and 14). The pre-peak feature is denoted by the arrow.

For the 8 nm PZT sample (Figure S11a), the $L_3$ peak positions are 641.49 eV, 641.46 eV, 641.46 eV and 641.39 eV from the bulk LSMO (number 1,2) to the Mn columns near the LSMO/PZT interface (number 13,14). The standard error of each fitting was used to calculate the error bars ($\pm$ 0.02 eV). For the 130 nm PZT sample (Figure S11b), the $L_3$ peak positions are 641.82 eV, 641.69 eV, 641.67 eV, and 641.49 eV from the bulk layer (number 1,2) to the top interface (number 13,14). The error bars are $\pm$ 0.04 eV.

In the case of LSMO with 8 nm PZT, the $L_3$ peak position difference between the bulk layer and the LSMO/PZT interface is small (~ 0.1 eV), which is possibly due to the error bar or the localized probe of EELS. The lower valence state in the interfacial LSMO than that of bulk thin film layer is revealed by the almost missing pre-peak feature in the O *K*-edge spectrum at the LSMO/PZT interface (Figure S11c) and the lowest valence state determined from the averaged EELS of Mn $L_{3,2}$-edge results (Figure 3b). In the case of LSMO with 130 nm PZT, the $L_3$ peak shifts to a lower energy position (~ 0.3 eV) from the bulk layer to the LSMO/PZT interface, suggesting a lower valence state in the interfacial LSMO than that of bulk thin film layer. The pre-peak feature, which still may be observed near the STO/LSMO interface, gradually vanishes at the LSMO/PZT interface (Figure S11d). These results are consistent with the proposed built-in electric field accompanied with the negative charge accumulation in the LSMO near the LSMO/PZT interface.

## S12. STEM-HAADF and the depth profile of out-of-plane lattice spacing

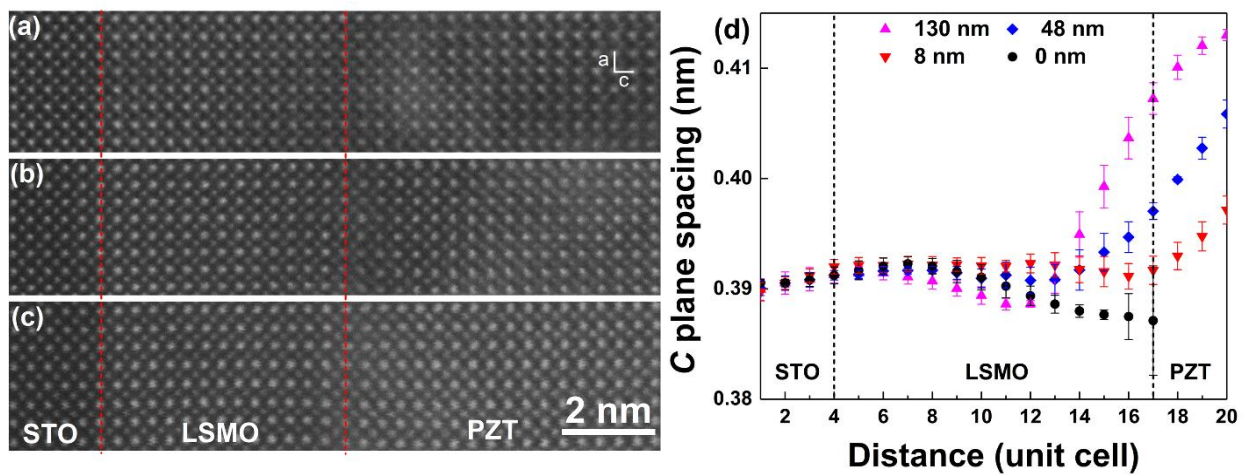

Figure S12: High angle annular dark field (HAADF) images in STEM mode for samples with the different PZT thickness: (a) 8 nm (b) 48 nm (c) 130 nm. The two red dash lines indicate the STO/LSMO and LSMO/PZT interface, respectively. (d) Out-of-plane lattice spacing as a function of distance from STO substrate to PZT. The dash lines correspond to the STO/LSMO and LSMO/PZT interface, respectively.

As shown in Figure S12a-S12c, the thickness of LSMO is kept at ~14 unit cells in all three samples. Close to the LSMO/PZT interface, there is no strain relaxation in PZT. Based on XRD results discussed above, the PZT is relaxed in a large area. Combining the XRD and STEM results, the gradual relaxation of strain in PZT contributes to the flexoelectric effect.

The change of out-of-plane lattice spacing as a function of distance from STO substrate to PZT was determined by fitting the positions of atoms from STEM-HAADF images. As shown in Figure S12d, the lattice spacing at LSMO/PZT interface is the largest for LSMO with 130 nm PZT and that is the smallest for LSMO with 8 nm PZT. In comparison, the pure LSMO sample shows no elongation of the lattice spacing near the surface. Therefore, the variation of local structure of interfacial LSMO is highly correlated with the thickness of overlying PZT film. In addition, the interfacial orbital occupancy of LSMO may be influenced by both the *c/a* ratio [13] and the orbital hybridization between Mn-O-Ti/Zr [14, 15]. The interfacial Mn $3d_{3z^2-r^2}$ orbital occupancy may be preferred when capped by 8 nm PZT in $P_{down}$ state due to a smaller out-of-plane lattice spacing and the enhanced orbital hybridization.

## S13. Half-integer diffraction results

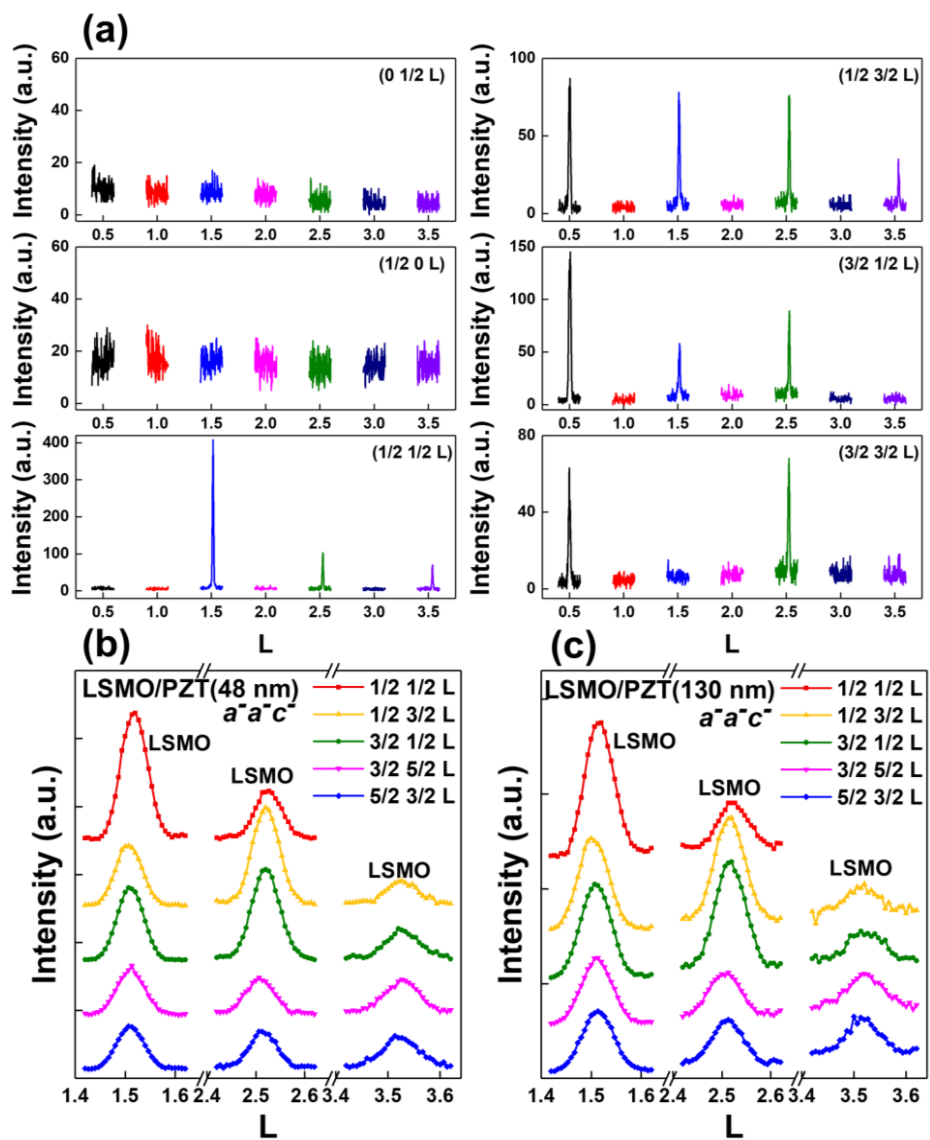

Figure S13: The half-integer diffraction peaks for LSMO films with (a) 0 nm (b) 48 nm (c) 130 nm PZT.

By the use of the synchrotron light source, a series of clear half-integer peaks are observed in LSMO with 0 nm, 48 nm, and 130 nm PZT. According to the relationship between the half-integer diffraction peaks and the type of tilts [16], we determined the octahedral rotation patterns of LSMO with different thicknesses of PZT films. Take pure LSMO sample for example (Figure S13a), (0, 0.5, L) and (0.5, 0, L) are used to check the presence of $a^+$ and $b^+$ type of tilt, respectively, and no such peaks are noted in pure LSMO sample. The disappearance of (0.5, 1.5, 1), (0.5, 1.5, 2), (1.5, 0.5, 1) and (1.5, 0.5, 2) peaks indicate no $c^+$ existing in pure LSMO sample. On the other hand, the presence of sharp peaks in (0.5, 0.5, L) and (1.5, 1.5, L) scans suggest the presence of $a^-/b^-$, and the sharp peaks in (0.5, 1.5, L) and (1.5, 0.5, L) scans indicate the presence of $a^-/b^-/c^-$. Due to the tetragonal structure of LSMO film with the same in-plane lattices, the tilt around the a and b axis should be equivalent. Therefore, the rotation pattern of pure LSMO film may be either $a^-a^-c^-$ or $a^-a^-c^0$. The experimental intensities were used to calculate the tilt angles and the oxygen positions of LSMO with 48 nm and 130 nm PZT. Our high-quality half-integer diffraction results may guarantee a reliable fitting result.

## S14. Dependence of deviation on octahedral tilt angles

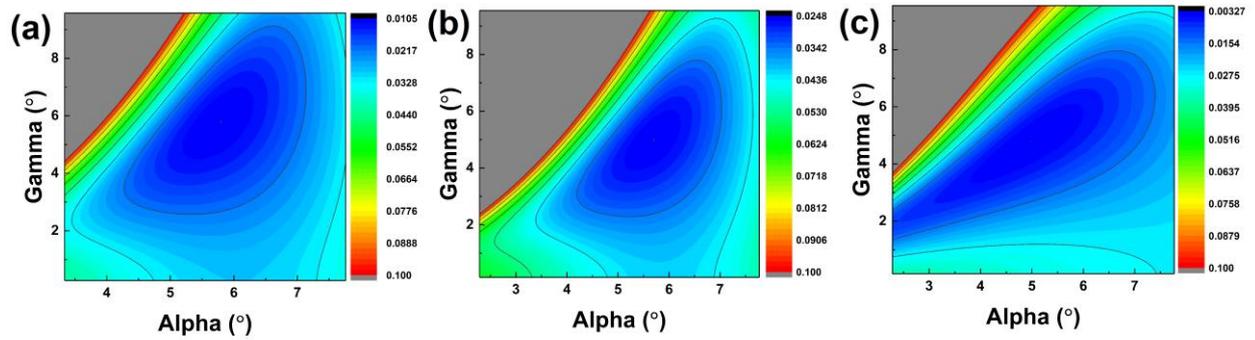

Figure S14: Dependence of deviation on octahedral tilt angles of STO/LSMO/PZT heterostructures with different thickness of PZT: (a) 8 nm (b) 48 nm and (c) 130 nm. In the fitting, the tilts around x, y axis are set to be the same ($α = β$) based on the tetragonal structure of LSMO with equivalent in-plane lattices ($a = b$).

There is one minimum point in the Figures S14a-S14c, which corresponds to the most reasonable fitting result. The tilt angles ($α, β, γ$) for LSMO with different thickness of PZT are listed in Table I of main text. The calculated and experimental intensity of half-integer peaks are summarized in Table SII.

## S15. ZFC and FC magnetization loops

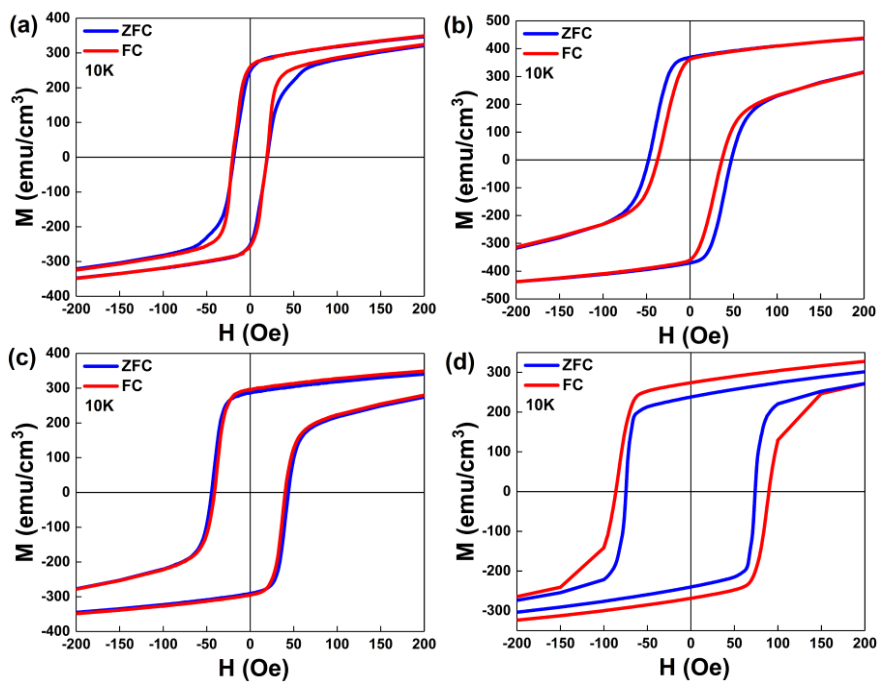

Figure S15: Measurements of zero field cooling (ZFC) and field cooling (FC) magnetization loops of (a) pure LSMO (b) LSMO/PZT (8 nm) (c) LSMO/PZT (48 nm) and (d) LSMO/PZT (130 nm) were conducted at T=10 K. The field cooling was applied with 5 T magnetic field. The measurement step size is 1 Oe.

As shown in Figures S15a-15c, no exchange bias and no increase of coercivity are found in the FC magnetization loops of pure LSMO, LSMO with 8 nm and 48 nm PZT films. In comparison, an obvious enhancement of coercivity with a small exchange bias (~1.7 Oe, $H_E = |H_{C1}+H_{C2}|/2$ where $H_{C1}$ and $H_{C2}$ are the forward and reversed coercive fields of magnetization hysteresis loops) is observed in the FC magnetization loop of LSMO with 130 nm PZT, as compared to that of the ZFC magnetization loop (Figure S15d). Considering the measurement step size is 1 Oe, this small value of exchange bias is trustworthy. In LSMO thin films (~14 u.c.), the thickness of antiferromagnetic (AFM) layers may be very small, leading to the weak AFM anisotropy [17]. The ferromagnetic layer would drag small AFM spins when it rotates during magnetization loop, resulting in an obvious increase in coercivity with a small loop shift. The presence of positive exchange bias (shift of the magnetic hysteresis loop in the same direction as the applied +5 T magnetic fields) may result from the antiferromagnetic coupling at interface between the FM and AFM layers [17, 18].

## S16. Summary of various techniques

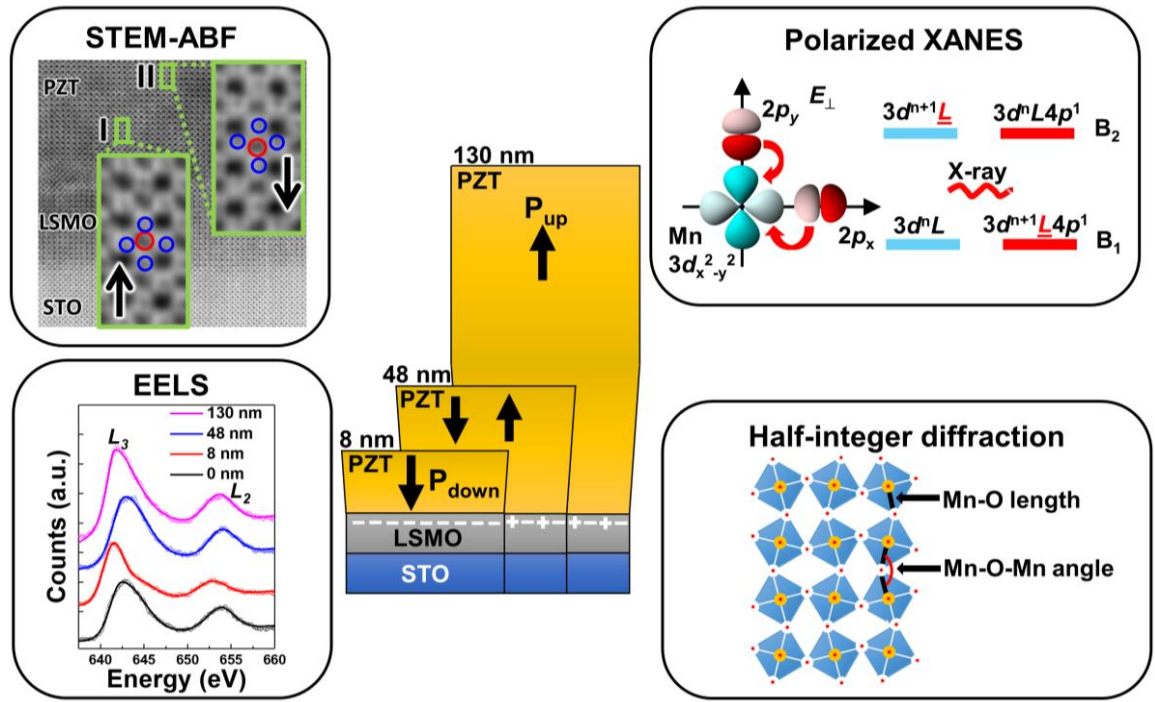

Figure S16: Schematic illustration of various measurements presented in the main text.

STEM-ABF was used to elucidate the self-polarization direction at the LSMO/PZT interface and the change of out-of-plane lattice constants as a function of distance from STO/LSMO to LSMO/PZT. EELS was used to probe the interfacial valence change of LSMO with different thicknesses of PZT films. Polarized XANES was utilized to demonstrate the anisotropic Mn-O charge transfer, while the half-integer diffraction measurements were explored to characterize the anisotropic local structures, including the Mn-O bond lengths and Mn-O-Mn bond angles. Combined results of polarized XANES and anisotropic bond lengths shed lights on the preferential orbital occupancy of LSMO.

**Table SI**

Table SI: $T_{MI}$ of LSMO films as a function of PZT thickness.

| PZT thickness (nm) | 0 | 4 | 8 | 16 | 48 | 96 | 130 |
|---|---|---|---|---|---|---|---|
| $T_{MI}$ (K) | 306 (± 3.06) | 305 (± 3.05) | 309 (± 3.09) | 303 (± 3.03) | 300 (± 3.00) | 296 (± 2.96) | 284 (± 2.84) |

$T_{MI}$ values were determined by the maximum of the first derivative of R-T curves, as shown in the inset of Figure 2d. The error bars of $T_{MI}$ were estimated based on the instrumental temperature accuracy (± 1%).

**Table SII**

Table SII: Summary of half-integer diffraction peaks of three samples from the experiment and the calculation.

| Peaks | LSMO/PZT (8 nm) | | Peaks | LSMO/PZT (48 nm) | | Peaks | LSMO/PZT (130 nm) | |
|---|---|---|---|---|---|---|---|---|
| | $I_{Exp.}$ | $I_{Cal.}$ | | $I_{Exp.}$ | $I_{Cal.}$ | | $I_{Exp.}$ | $I_{Cal.}$ |
| 0.5 0.5 1.5 | 1 | 1 | 0.5 0.5 1.5 | 1 | 1 | 1.5 0.5 2.5 | 1 | 1 |
| 0.5 0.5 2.5 | 0.37 | 0.33 | 0.5 0.5 2.5 | 0.40 | 0.33 | 0.5 0.5 2.5 | 0.43 | 0.49 |
| 0.5 1.5 1.5 | 0.51 | 0.48 | 0.5 1.5 1.5 | 0.54 | 0.44 | -0.5 -0.5 2.5 | 0.52 | 0.49 |
| 0.5 1.5 2.5 | 0.75 | 0.68 | 0.5 1.5 2.5 | 0.86 | 0.68 | -0.5 0.5 2.5 | 0.40 | 0.49 |
| 0.5 1.5 3.5 | 0.25 | 0.24 | 0.5 1.5 3.5 | 0.26 | 0.25 | 0.5 1.5 1.5 | 0.70 | 0.69 |
| 1.5 0.5 1.5 | 0.62 | 0.48 | 1.5 0.5 1.5 | 0.62 | 0.44 | 0.5 1.5 2.5 | 0.95 | 1 |
| 1.5 0.5 2.5 | 0.74 | 0.68 | 1.5 0.5 2.5 | 0.82 | 0.68 | 1.5 0.5 1.5 | 0.72 | 0.69 |
| 1.5 0.5 3.5 | 0.29 | 0.24 | 1.5 0.5 3.5 | 0.31 | 0.25 | 1.5 2.5 2.5 | 0.45 | 0.47 |
| 1.5 2.5 1.5 | 0.44 | 0.65 | 1.5 2.5 1.5 | 0.43 | 0.6 | 2.5 1.5 2.5 | 0.49 | 0.47 |
| 1.5 2.5 2.5 | 0.29 | 0.28 | 1.5 2.5 2.5 | 0.37 | 0.28 | -1.5 0.5 1.5 | 0.62 | 0.69 |
| 1.5 2.5 3.5 | 0.34 | 0.35 | 1.5 2.5 3.5 | 0.39 | 0.37 | -1.5 0.5 2.5 | 0.95 | 1 |
| 2.5 1.5 2.5 | 0.34 | 0.28 | 2.5 1.5 1.5 | 0.41 | 0.60 | 1 0.5 1.5 | 0 | 0 |
| 2.5 1.5 3.5 | 0.34 | 0.35 | 2.5 1.5 2.5 | 0.38 | 0.28 | 1 0.5 2 | 0 | 0 |
| 2.5 2.5 3.5 | 0.20 | 0.33 | 2.5 1.5 3.5 | 0.35 | 0.37 | 1 0.5 2.5 | 0 | 0 |
| 0.5 -0.5 1.5 | 0.98 | 1 | 2.5 2.5 1.5 | 0.15 | 0.31 | 1.5 1 1.5 | 0 | 0 |
| -0.5 0.5 1.5 | 0.90 | 1 | 2.5 2.5 3.5 | 0.25 | 0.37 | 1.5 1 2 | 0 | 0 |
| -0.5 -0.5 1.5 | 0.99 | 1 | -0.5 0.5 1.5 | 0.98 | 1 | 1.5 1 2.5 | 0 | 0 |
| 1 0.5 1.5 | 0 | 0 | 0.5 -0.5 1.5 | 1 | 1 | | | |
| 1 0.5 2 | 0 | 0 | -0.5 -0.5 1.5 | 1 | 1 | | | |
| 1 0.5 2.5 | 0 | 0 | 1 0.5 1.5 | 0 | 0 | | | |
| 1.5 1 1.5 | 0 | 0 | 1 0.5 2 | 0 | 0 | | | |
| 1.5 1 2 | 0 | 0 | 1 0.5 2.5 | 0 | 0 | | | |
| 1.5 1 2.5 | 0 | 0 | 1.5 1 1.5 | 0 | 0 | | | |
| | | | 1.5 1 2 | 0 | 0 | | | |
| | | | 1.5 1 2.5 | 0 | 0 | | | |

The experimental intensities of half-integer peaks extracted from Figures 4b and Figures S13b-S13c are summarized in Table SII. These experimental intensities $I_{Exp.}$ were used to calculate the tilt angles and the position of oxygen atoms by the home-made codes [19, 20]. The intensity fitting is based on the following equation [21]:

$$I = I_0 \frac{1}{\sin(\eta)} \frac{1}{\sin(2\theta)} (\sum_{j=1}^{4} D_j |F_{hkl}|^2)$$

where $I_0$ is the incident x-ray flux, $\eta$ is the incident x-ray angle, $2\theta$ is the scattering x-ray angle, $F_{hkl}$ is the structure factor for the oxygen atoms, and $D_j$ is the relative volume of structural domains. The structural domains may be interpreted as different possible atomic oxygen spatial arrangement within one octahedron. In our study, the relative volume of structural domain $D_j$ were acquired from a set of symmetrically equivalent half-integer scans with a fixed $L$. The equal domain populations were determined in all three samples. The structure factor $F_{hkl}$ can be described as follows:

$$F_{hkl} = f_{O^{2-}} \sum_{n=1}^{24} \exp[2\pi i (hu_n + kv_n + lw_n)]$$

where $f_{O^{2-}}$ is the form factor [22] and $(u_n, v_n, w_n)$ is the position of $n$ th oxygen within the unit cell. By fitting the $I_{Exp.}$, the oxygen atomic coordinates determined by the tilt angles ($\alpha$, $\beta$, $\gamma$) along the three axes were acquired with the most reasonable fitting results (i.e. the smallest deviation ($\sigma = \sum_{n=1}^{i}(I_{exp,n} - I_{cal,n})^2$), as shown in Figures S14a-S14c. The more the number of half-integer peaks were used for the calculation, the more accurate fitting results will be acquired. By use of the synchrotron light source in APS, a number of half-integer peaks with strong intensities are observed in all three sample. Hence, the fitting results of the tilt angles are reliable based on the high-quality experimental data. After the tilt angles have been determined, the atomic coordinates for all atoms inside the unit cell could be calculated. By drawing the atomic arrangement in software "Diamond", the Mn-O bond length and Mn-O-Mn bond angle were measured.